\documentclass[traditabstract]{aa} 

\usepackage{graphicx}
\usepackage{natbib}
\usepackage{lscape}
\usepackage{longtable}
\usepackage{txfonts}
\usepackage{caption}
\usepackage{subcaption}

\usepackage{float}

\def\mstar  {$M_{\star}$}
\def\macc   {$\dot{M}_{\rm acc}$}
\def\lacc   {$L_{\rm acc}$}

\def\msun {$M_{\odot}$}
\def\lsun {$L_{\odot}$}
\def\lstar {$L_\star$}

\def\lline {$L_{\rm line}$}

\defcitealias{Herczeg14}{HH14}

\usepackage{color}

\begin{document}

   \title{X-Shooter study of accretion in Chamaeleon I\thanks{This work is based on observations made with ESO Telescopes at the Paranal Observatory under programme ID 084.C-1095 and 094.C-0913. }}

   \author{C.F. Manara \inst{1}\fnmsep\thanks{ESA Research Fellow}, D. Fedele\inst{2,3}, G.J. Herczeg\inst{4},
          \and
        P.S. Teixeira\inst{5}
          }

   \institute{Scientific Support Office, Directorate of Science and Robotic Exploration, European Space Research and Technology Centre (ESA/ESTEC), Keplerlaan 1, 2201 AZ Noordwijk, The Netherlands \\
              \email{cmanara@cosmos.esa.int}
         \and
        INAF-Osservatorio Astroﬁsico di Arcetri, L.go E. Fermi 5, I-50125 Firenze, Italy
         \and
             Max-Planck Institut f\"ur Extraterrestrische Physik, Giessenbachstrasse 1, 85748 Garching, Germany
         \and
             Kavli Institute for Astronomy and Astrophysics, Peking University, Yi He Yuan Lu 5, Haidian Qu, Beijing 100871, China
        \and
        Universtit\"at Wien, Institut f\"ur Astrophysik, T\"urkenschanzstrasse 17, 1180 Vienna, Austria
             }

   \date{Received August 21, 2015; accepted October, 23, 2015}

 
\abstract{We present the analysis of 34 new VLT/X-Shooter spectra of young stellar objects in the Chamaeleon~I star-forming region, together with four more spectra of stars in Taurus and two in Chamaeleon~II. The broad wavelength coverage and accurate flux calibration of our spectra allow us to estimate stellar and accretion parameters for our targets by fitting the photospheric and accretion continuum emission from the Balmer continuum down to $\sim$700 nm. The dependence of accretion on stellar properties for this sample is consistent with previous results from the literature. The accretion rates for transitional disks are consistent with those of full disks in the same region. The spread of mass accretion rates at any given stellar mass is found to be smaller than in many studies, but is larger than that derived in the Lupus clouds using similar data and techniques. Differences in the stellar mass range and in the environmental conditions between our sample and that of Lupus may account for the discrepancy in scatter between Chamaeleon~I and Lupus.
Complete samples in Chamaeleon I and Lupus are needed to determine whether the difference in scatter of accretion rates and the lack of evolutionary trends are not influenced by sample selection. 
}



   \maketitle
%

\section{Introduction}\label{sect::introduction}

During the first Myr of evolution toward the main sequence, the disk around young stellar objects (YSOs) evolves and is dispersed mainly by disk accretion processes and photoevaporation \citep{AlexanderPPVI}. The importance and the properties of disk accretion are constrained by studying objects located in different star-forming regions and at different evolutionary phases. 

Observations of accretion in YSOs mostly focus on determining the amount of material accreted onto the central star per unit time. This mass accretion rate \citep[\macc, e.g.,][]{Hartmann98} may be compared to the stellar and disk properties to test the predictions of models of disk evolution. Relations between the accretion luminosity (\lacc) and stellar luminosity \citep[\lstar, e.g.,][]{Natta06,Clarke06,Rigliaco11a,Manara12} and between \macc \ and stellar mass \citep[\mstar, e.g.,][]{Muzerolle03,Mohanty05,Natta06,Herczeg08,Antoniucci11,Antoniucci14,Manara12,Alcala14,Ercolano14,Frasca15} are frequently derived to examine the dependence of accretion on the properties of the central star.
Accretion has sometimes been found to correlate with age \citep[e.g.,][]{Hartmann98,Sicilia-Aguilar10,Antoniucci14}, although ages of individual young stars in a given star-forming region are very uncertain as a result of both observational and model-dependent uncertainties \citep[e.g.,][]{SoderblomPPVI}, including a dependence of age estimate on the spectral type \citep[e.g.,][]{Herczeg15}. 

Measurements of \macc \ are obtained spectroscopically using direct or indirect methods. 
Flux-calibrated spectra that span a wide wavelength range and include the Balmer continuum region yield direct estimates of the \lacc \ while also allowing for the incorporation of veiling into spectral type and extinction measurements.  This approach is thought to lead to more reliable estimates for the stellar mass and radius, which are necessary ingredients when estimating \macc \ and ages (\citealt{Manara13b}; \citealt[][hereafter HH14]{Herczeg14}).
While the entire UV spectrum is obtained only from space telescopes and is then used to derive \macc \ for relatively small samples of YSOs \citep[e.g.,][]{Gullbring00,Ingleby13}, recent instruments mounted on very large telescopes give the community access to large quantity of spectra in the region of the visible UV from Earth, namely from $\sim$300 nm to $\sim$400 nm, which is sufficient for capturing the Balmer jump. Modeling the excess continuum emission at these wavelengths yields direct measurements of \lacc. This method has been used in the past on a substantial number of objects \citep[e.g.,][]{Gullbring98,Herczeg08,Ingleby14}, but often using non-simultaneous and low-resolution spectra at optical wavelengths to determine the photospheric properties of the target. On the other hand, the indirect method of measuring the luminosity of various emission lines (\lline) and converting this into \lacc \ through calibrated \lacc-\lline \ relations \citep[e.g.,][]{Muzerolle98,Fang09} can be efficiently used on spectra covering a relatively narrow wavelength range, such as those offered by multi-object spectrographs. Recent studies have also demonstrated that \lacc \ derived from several emission lines measured simultaneously lead to results compatible, although with a larger scatter, with more direct methods \citep{Rigliaco12,Alcala14}. Therefore, measuring \macc \ from the luminosity of emission lines is a method that is being extensively used in the literature \citep[e.g.,][]{Mohanty05,Natta06,Fang09,Costigan12, Antoniucci11,Antoniucci14,Biazzo14}. 

The second-generation VLT instrument X-Shooter allows us to obtain simultaneous spectra of the whole wavelength range from $\sim$300 nm to $\sim$2500 nm at medium resolution and to thus derive \lacc \ directly from the UV-excess and stellar properties from the photospheric features of the spectra. \citet{Manara13b} have developed a new method to model all the components of the observed X-Shooter spectra of YSOs to directly derive \lacc \ simultaneously
with the stellar parameters and extinction. First studies of accretion with this instrument focused on individual targets \citep[e.g.,][]{Rigliaco11b,Stelzer13a} or on a few very low-mass stars and BDs in $\sigma$-Orionis \citep{Rigliaco12}. The largest sample studied to date with this instrument is a set of 36 YSOs in the Lupus I and III clouds \citep{Alcala14}. This study found a significantly smaller spread of values of \macc \ at any given \mstar \ with respect to previous works. This result opened new questions in the field, such as whether Lupus is a special region where the initial conditions of star formation are similar for different objects, or if the instrinsic YSO variability in this region is inherently small. At the same time, the narrow \macc \ distribution derived using X-Shooter spectra started to show the advantage of high-quality and broad-band spectra in obtaining more precise accretion rates. A large sample of 22 transitional disks \citep{Manara14} has also been studied with this instrument and the method by \citet{Manara13b}, but larger samples of known full disks in various star-forming regions studied with the same method are needed to determine the general evolution of accretion in YSOs. 

In this paper we present a study of a large sample of 34 Class~II YSOs in the Chamaeleon~I star-forming region \citep[$d$ = 160 pc,][]{LuhmanCha} observed with X-Shooter, and of a few additional targets in Taurus and Chamaeleon~II. These results are used to verify whether the small spread of \macc \ found in Lupus is confirmed in another star-forming region and to extend the comparison between accretion and stellar properties to solar-mass YSOs, which were not targeted by \citet{Alcala14}. 
The paper is organized as follows. First, the observations and data reduction procedures are explained in Sect.~\ref{sect::obs}, then the method used to derive stellar and accretion parameters is presented and compared with literature estimates in Sect.~\ref{sect::methodsect}. The results are then presented in Sect.~\ref{sect::results} and are discussed in Sect.~\ref{sect::discussion}, including a comparison with other studies with similar methods in other star-forming regions. Finally, Sect.~\ref{sect::conclusions} presents the main conclusions of this work.

%


\section{Observations and data reduction}\label{sect::obs}

\subsection{Main sample}
The sample of pre-main sequence (PMS) stars presented in this study was selected from \citet{Luhman04,Luhman07}, \citet{Spezzi08}, and \citet{Alcala08} and has been observed with
X-Shooter \citep{Vernet11} on the Very Large Telescope (Paranal,
Chile) between January 17 and 19, 2010. X-Shooter provides simultaneous
medium-resolution ($R \sim$ 5000 - 18000, depending on the spectral arm and the slit width) spectroscopic observations
between 300-2500\,nm. It consists of three independent cross-dispersed
echelle spectrographs, referred to as the UVB, VIS, and NIR arms, with the following spectral coverage: UVB
(300-559.5\,nm), VIS (559.5-1024\,nm), and NIR (1024-2480\,nm). The
observations were executed by nodding the telescope along a direction
perpendicular to the slit. Before the nodding observation, each
target was observed in stare mode using a large slit width of
1\farcs5 - 5\arcsec. Large-slit spectra have been used to properly flux calibrate the spectra obtained with narrow slits, which are used for the analysis because they lead to a higher spectral resolution. 
The observing log is reported in Table~\ref{tab::log} and includes the time of observation, the exposure times, the number of nodding positions, and the width of the slits used for the high-resolution observations.

\smallskip
\noindent
The spectra were reduced with the ESO X-Shooter pipeline \citep{Modigliani10} version 
v1.3.2 following a standard reduction scheme: bias subtraction 
(stare mode only), flat-fielding, wavelength calibration, background removal 
(stare mode only), and spectrum extraction. The latter is performed manually on the rectified spectrum, and after the extraction of each component of a binary system when both were included in the slit and spatially separated. For the spectra taken in the nodding mode, the background removal is obtained by the difference of spectra pairs.
The UVB and VIS arms are corrected for atmospheric extinction using the atmospheric transmission curve of Paranal by \citet{Patat11}.

\smallskip
\noindent
The spectrophotometric standard star GD-71 was observed at the beginning of each night.
The spectral response function of the instrument, which gives the conversions from ADU 
to erg\,cm$^{-2}$\,s$^{-1}$\,\AA$^{-1}$, was obtained by dividing the observed spectrum by the tabulated 
spectrum provided with the X-Shooter pipeline. This has been derived from observations of GD-71 
for each night and each arm independently. The response 
function for the UVB and VIS\ arms does not show variations during the three nights of observation. Small variations 
are observed for the NIR arm, and the response function is taken as the average
of the three nights. The overall flux calibration accuracy was tested on telluric standard star observations and is stable with a $\sim$2\% accuracy.
Finally, the narrow-slit spectra are scaled upward to match the large-slit spectra. The scaling 
factor is computed as the ratio of the spectral continuum of the two spectra. More information on the reduction and flux calibration of the spectra obtained in this program will be discussed in Rugel et al., in prep.

In total, 40 bona-fide Class~II members of the Chamaeleon~I star-forming region were observed. Here we discuss the properties of 34 of these objects. The other six objects are transitional disks sources that were analyzed by \citet{Manara14} with the same method as
was implemented in this paper, and they are included in our analysis. The membership of 22 YSOs in the Chamaeleon~I sample has been confirmed by recent proper motion studies \citep{Lopez-Marti13}. 
Following \citet[][and references therein]{LuhmanCha}, the adopted distances to objects in the Chamaeleon~I region is 160 pc. 
In addition to these targets, two objects located in the Chamaeleon~II cloud \citep[$d$=178 pc,][]{LuhmanCha}, both confirmed members according to \citet{Lopez-Marti13}, and four targets located in the Taurus molecular cloud \citepalias[$d$=131-140 pc,][]{Herczeg14} were observed, and their accretion properties are reported here. These targets were observed with the same observing strategy as the Chamaeleon~I objects and thus are interesting objects
with which to compare the accretion properties of YSOs in different star-forming regions.

\subsection{Additional templates used for the analysis}\label{sect::classIII}

The analysis presented here makes use of several photospheric templates observed with X-Shooter to estimate the stellar parameters of the targets. These templates are non-accreting Class~III YSOs collected and characterized mainly by \citet{Manara13a}. This set of templates is almost complete for objects with SpT in the M-subclass, but includes only few objects with SpT in the K-subclass (one K5, two K7 YSOs). Thus, it is useful to include some additional templates here. Three of them have been discussed in \citet{Manara14} and have SpT G4, G5, and K2. Finally, another target, HBC407, is included here. This object is a Class~III YSO \citep{Luhman10} with SpT K0 \citepalias{Herczeg14}, which was observed during the program Pr.Id.094.C-0913 (PI Manara). Here we discuss the observations, data reduction, and classification for this target.

HBC407 was observed in service mode during the night of December
8, 2014 with the following strategy. First, a complete nodding cycle ABBA was performed using the narrower slits 0.5\arcsec-0.4\arcsec-0.4\arcsec in the three arms, respectively, with an exposure time for each exposure of 430 s in the UVB arm, 375 s in the VIS arm, and three subintegrations of 125s each in the NIR arm. Then, the object was observed immediately after this with the larger slits (5.0\arcsec) with an AB nodding cycle with single exposures of 45 s in the UVB and VIS arms and three times 15 s in the NIR arm. The latter was used to obtain a proper flux calibration of the spectra without slit losses. This spectrum was reduced using the X-Shooter pipeline version 2.5.2, and flux calibration of the spectra was made by the pipeline using the standard star FEIGE-110. We then manually rescaled the spectra obtained with the narrow slits to those obtained with the large slits and found that they agree very well with the available photometry. Telluric correction was carried out using the telluric standard Hip036392 and the procedure described by \citet{Alcala14}. 

This object has been classified as a K0 YSO with extinction $A_V$=0.8 mag by \citetalias{Herczeg14}. Using the spectral indices discussed in that work, in particular using the R5150 index, we obtain the same SpT for this target. We then compared the observed spectrum with a BT-Settl AGSS2009 synthetic spectrum \citep{Allard11} at $T_{\rm eff}$=5100 K and log$g$=-4.0 and found a best agreement by applying a reddening correction of $A_V$=0.8 mag using the extinction law by \citet{Cardelli} and $R_V$=3.1. This $T_{\rm eff}$ is consistent with a SpT between G8 and K1 according to different SpT-$T_{\rm eff}$ relations (\citetalias{Herczeg14} and \citet{KH95}, respectively). Here a SpT K0 is assumed for this target. Finally, we calculated the luminosity of the target from the observed spectrum and the synthetic spectrum following the same procedure as \citet{Manara13a}, and, assuming $d$=140 pc \citepalias{Herczeg14}, we derive log(\lstar/\lsun) = $-$0.45 for this target, the same value as \citetalias{Herczeg14}. However, the position of this target is below the main sequence. This could suggest that it is not a member of the Taurus region. Nevertheless, this uncertainty does not influence the estimates of stellar parameters for other targets based on this template because the information on the distance is taken into account in the procedure.

The spectrum of this target shows no signs of accretion in the typical lines tracing accretion in YSOs (e.g., H$\alpha$), but a prominent LiI~$\lambda$670.8 nm absorption line, and narrow chromospheric emission lines as a reversal in the core of the photospheric absorption lines in the Ca IR triplet. 


\section{Analysis method}\label{sect::methodsect}

Here we briefly explain our method to self-consistently derive stellar and accretion parameters from
the X-Shooter spectra, and we report our results. All the objects discussed here have also been previously characterized by other authors \citep[e.g.,][see Table~\ref{tab::lit}]{Luhman07} using red optical and infrared spectra. The comparison with these results is also presented in this section. 

\subsection{Determining the stellar and accretion parameters}\label{sect::method}
To determine the stellar and accretion parameters, we used the method described in \citet{Manara13b}, which determines the SpT, $A_V$, \lstar, and \lacc \ by finding the best fit among a grid of models. This grid includes different photospheric templates (see Sect.~\ref{sect::classIII}), several slab models to reproduce the excess continuum emission due to accretion, and extinction parametrized with the reddening law by \citet{Cardelli} with $R_V$=3.1 and values of $A_V$ from 0 mag to 10 mag with steps of 0.1 mag.
The slab model has been used by \citet{Valenti93} and \citet{Herczeg08}, for example, and it is not a physical model of the accretion region, unlike the shock models by \citet{Calvet98}, but provides an accurate bolometric correction to determine the total excess flux that is due to accretion. The best fit is found by comparing the flux of the observed spectrum and that of each model in various continuum regions and in different features of the spectra, and by minimizing a $\chi^2_{\rm like}$ distribution to determine the best match. The points included in the fitting procedure range from $\lambda\sim$330 nm to $\lambda\sim$715 nm. A proper fit of the Balmer continuum is needed to determine \lacc \ and the veiling of the spectra. The Balmer continuum is well determined even with spectra with very low signal-to-noise ratio (S/N) on the continuum at  $\lambda\sim$330 nm. The spectra of most of the targets discussed here have enough signal to perform the analysis. Only spectra with no signal in the Balmer continuum regions, such as those of Cha H$\alpha$ 9 and Cha H$\alpha$ 1 in this work, are not suitable for the method described here. More discussion of these and other individual targets is available in Appendix~\ref{sect::ind_targ}. The stellar parameters ($T_{\rm eff}$, \lstar) are derived from the properties of the best-fit template and \lacc \ from the slab model parameters as described by \citet{Manara13b}, and are reported in Table~\ref{tab::macc}. The best fit of the UV excess for all the targets are shown in Fig.~\ref{fig:best_fits}-\ref{fig::best_fits4}. We have compared our direct measurements of \lacc \ from the excess Balmer continuum emission to alternative estimates from line luminosities in the same spectrum. The agreement is good when using the \lacc-\lline \ relations by \citet{Alcala14}, with only small differences for the objects with SpT K2. For further analyses of the emission lines present in the spectra, we refer to Fedele et al. (in prep.).

\begin{figure}[!t]
\centering
\includegraphics[width=0.5\textwidth]{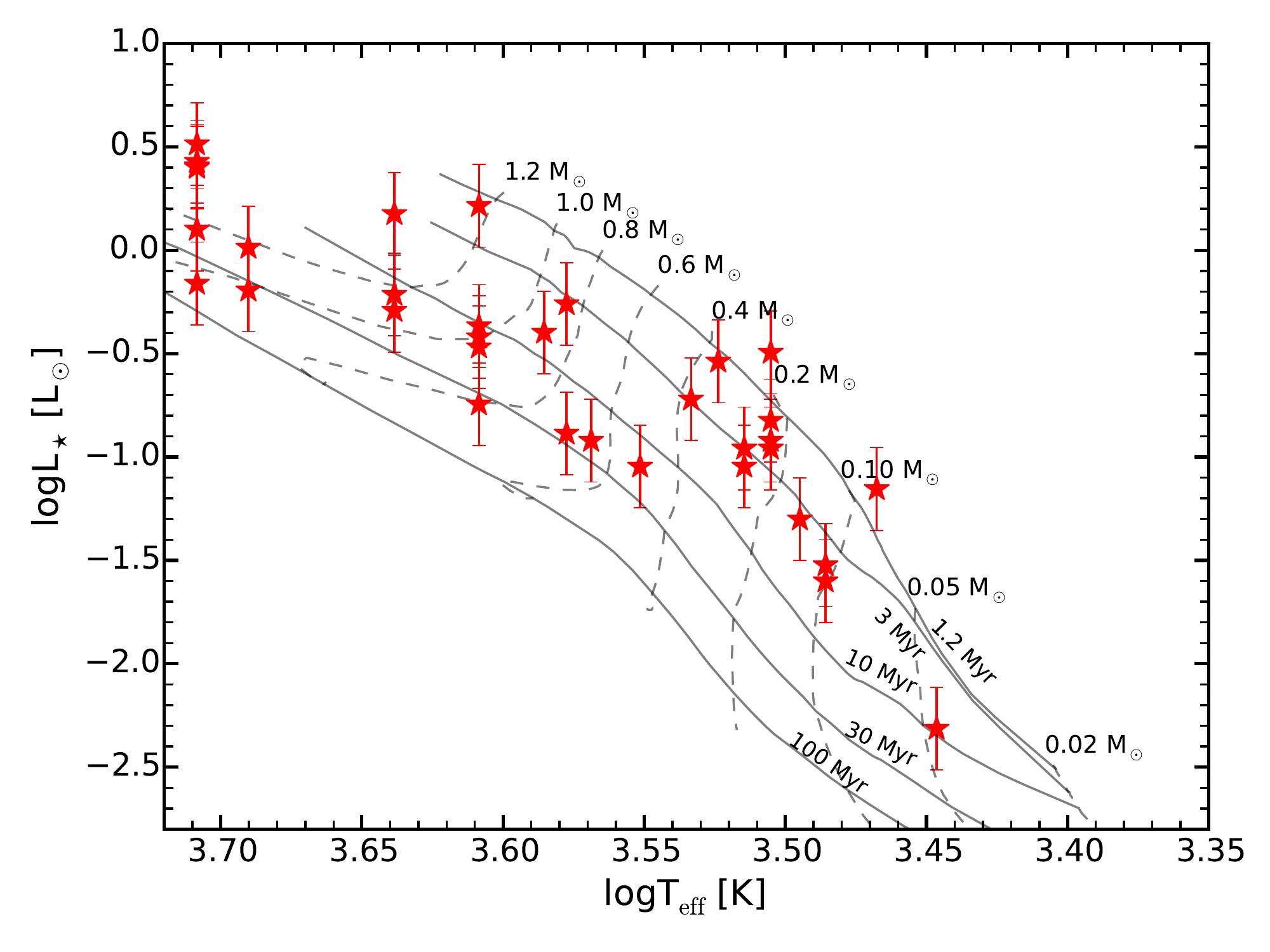}
\caption{H-R diagram of the Chamaeleon~I objects discussed here overplotted on evolutionary models from \citet{Baraffe98}.
     \label{fig::hrd}}
\end{figure}

The HR diagram (HRD) of the Chamaeleon~I sample is shown in Fig.~\ref{fig::hrd} overplotted on the isochrones from the evolutionary models by \citet{Baraffe98}. This evolutionary model was chosen for consistency with previous analyses of X-Shooter spectra, so that the comparison is less model dependent. Most of the objects are located on this diagram around the 3 Myr isochrone, with a few of objects located closer to 10 Myr, and only three objects are located slightly below the 30 Myr isochrone, but still above the 30 Myr isochrone from other models \citep[e.g.,][]{Siess00}. 
The stellar mass (\mstar)  was obtained by comparing the position in the HRD to the \citet{Baraffe98} evolutionary tracks. Finally, the stellar radius ($R_\star$) was derived from $T_{\rm eff}$ and \lstar, and \macc \ from the usual relation \macc=1.25$\cdot$\lacc$R_\star$/(G\mstar) \citep[e.g.,][]{Gullbring98}.

\begin{figure}[!t]
\centering
\includegraphics[width=0.5\textwidth]{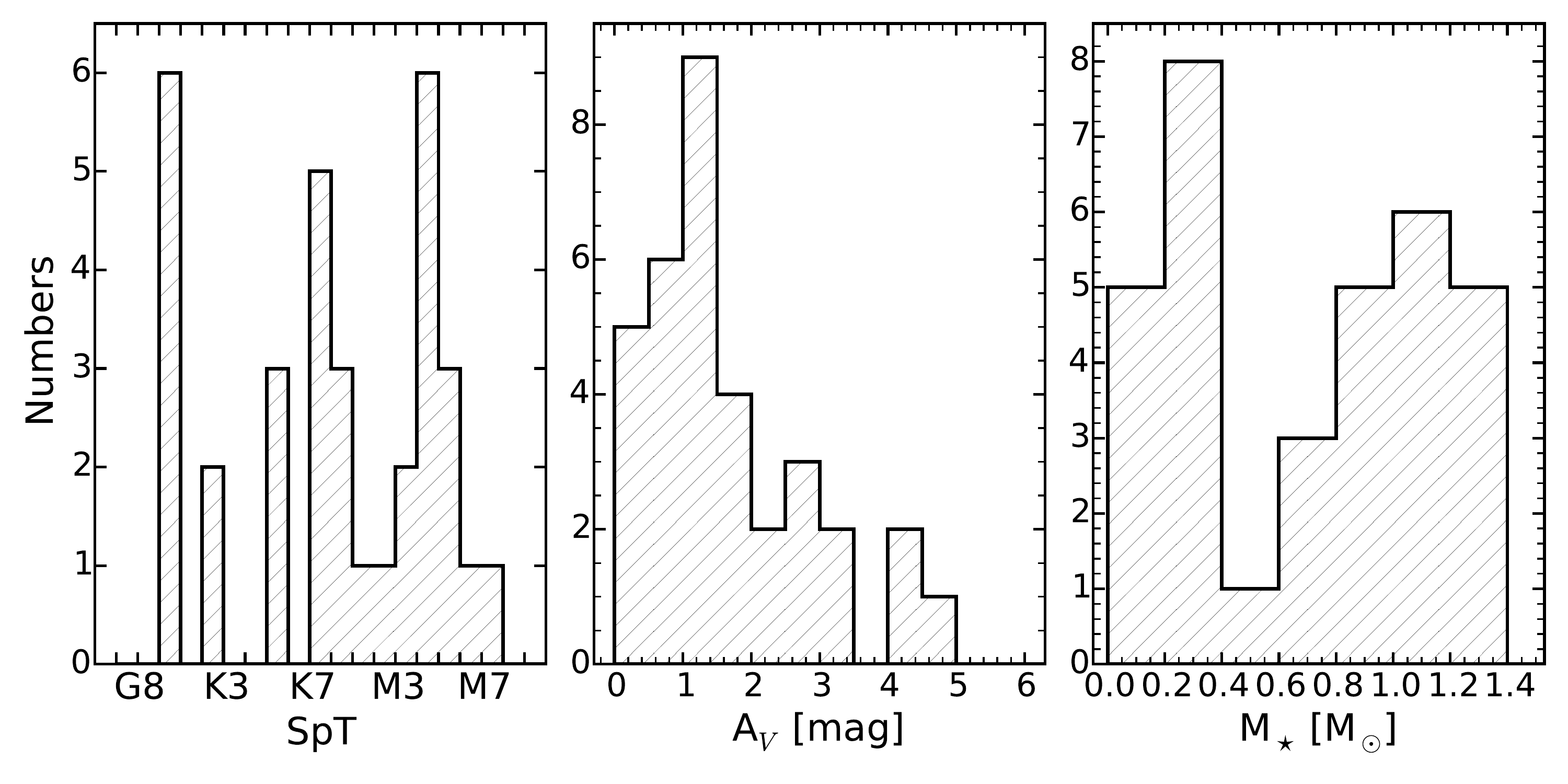}
\caption{Histogram of stellar properties (SpT, $A_V$, and \mstar) for the Chamaeleon~I sample.
     \label{fig::hist_pars}}
\end{figure}

The distribution of the derived parameters SpT, $A_V$, and \mstar \ for the Chamaeleon~I targets is shown in Fig.~\ref{fig::hist_pars}. As expected, most object have low $A_V\lesssim$3 mag. The Chamaeleon~I sample discussed here, together with the six transitional disks in Chamaeleon~I discussed by \citet{Manara14}, includes more than 40\% of the known Class~II and transitional disks YSOs in the region, with higher completeness levels at SpT earlier than M3, corresponding to \mstar$\sim$0.5 \msun, than at later SpT. This sample is thus large enough to discuss general properties of YSOs with \mstar$\gtrsim$0.5 \msun \ in the Chamaeleon~I region alone.

\begin{table*}  
\begin{center} 
\footnotesize 
\caption{\label{tab::macc} Derived stellar and accretion parameters for the Chamaeleon~I sample } 
\begin{tabular}{l| ccc| cc | cc } 
\hline \hline 
Object &  SpT & T$_{\rm eff}$ & A$_V$ & \lstar & log\lacc & \mstar & log\macc \\  
      &     \hbox{} & [K] & [mag] & [\lsun] & [\lsun] & [\msun] & [\msun/yr]   \\       
\hline 
T3 & K7 & 4060 & 2.6 & 0.18 & -1.25 & 0.79 & -8.62 \\ 
 T3-B & M3 & 3415 & 1.3 & 0.19 & -1.66 & 0.37 & -8.53 \\ 
 T4$^\dag$ & K7 & 4060 & 0.5 & 0.43 & $<$-2.24 & 1.03 & $<$-9.53 \\ 
 TW Cha & K7 & 4060 & 0.8 & 0.38 & -1.66 & 1.00 & -8.96 \\ 
 CR Cha$^*$ & K0 & 5110 & 1.3 & 3.26 & -1.42 & 1.78 & -8.71 \\ 
 T12 & M4.5 & 3200 & 0.8 & 0.15 & -2.12 & 0.23 & -8.77 \\ 
 CT Cha A & K5 & 4350 & 2.4 & 1.50 & 0.37 & 1.40 & -6.85 \\ 
 ISO-ChaI 52$^\dag$ & M4 & 3270 & 1.2 & 0.09 & $<$-3.79 & 0.25 & $<$-10.63 \\ 
 Hn5 & M5 & 3125 & 0.0 & 0.05 & -2.56 & 0.16 & -9.27 \\ 
 T23 & M4.5 & 3200 & 1.7 & 0.32 & -1.65 & 0.33 & -8.29 \\ 
 T24 & M0 & 3850 & 1.5 & 0.40 & -1.48 & 0.91 & -8.68 \\ 
 Cha H$\alpha$1 & M7.5 & 2795 & 0.0 & 0.00 & -5.11 & 0.05 & -11.69 \\ 
 Cha H$\alpha$9 & M5.5 & 3060 & 4.8 & 0.03 & -4.19 & 0.10 & -10.85 \\ 
 Sz22 & K5 & 4350 & 3.2 & 0.51 & -1.03 & 1.09 & -8.37 \\ 
 VW Cha & K7 & 4060 & 1.9 & 1.64 & -0.78 & 1.24 & -7.86 \\ 
 ESO H$\alpha$562 & M1 & 3705 & 3.4 & 0.12 & -2.01 & 0.66 & -9.32 \\ 
 T33-A$^\dag$ & K0 & 5110 & 2.5 & 1.26 & $<$-1.62 & 1.15 & $<$-9.93 \\ 
 T33-B & K0 & 5110 & 2.7 & 0.69 & -1.32 & 0.95 & -8.67 \\ 
 ISO-ChaI 143 & M5.5 & 3060 & 1.3 & 0.03 & -3.38 & 0.11 & -10.02 \\ 
 Cha H$\alpha$6 & M6.5 & 2935 & 0.1 & 0.07 & -3.86 & 0.10 & -10.25 \\ 
 T38 & M0.5 & 3780 & 1.9 & 0.13 & -2.02 & 0.71 & -9.35 \\ 
 T40 & M0.5 & 3780 & 1.2 & 0.55 & -0.48 & 0.87 & -7.61 \\ 
 T44 & K0 & 5110 & 4.1 & 2.68 & 0.62 & 1.40 & -6.60 \\ 
 T45a$^\dag$ & K7 & 4060 & 1.1 & 0.34 & $<$-2.59 & 0.97 & $<$-9.91 \\ 
 ISO-ChaI 237 & K5 & 4350 & 4.1 & 0.61 & -2.47 & 1.15 & -9.79 \\ 
 T49 & M3.5 & 3340 & 1.0 & 0.29 & -0.81 & 0.36 & -7.56 \\ 
 CHX18N & K2 & 4900 & 0.8 & 1.03 & -0.74 & 1.17 & -8.06 \\ 
 T51 & K2 & 4900 & 0.1 & 0.64 & -0.79 & 0.97 & -8.13 \\ 
 T51-B & M2 & 3560 & 0.5 & 0.09 & -1.92 & 0.51 & -9.12 \\ 
 T52 & K0 & 5110 & 1.0 & 2.55 & -0.19 & 1.40 & -7.42 \\ 
 T54-A$^\dag$ & K0 & 5110 & 1.2 & 2.51 & $<$-2.29 & 1.40 & $<$-9.53 \\ 
 Hn17$^\dag$ & M4.5 & 3200 & 0.4 & 0.11 & $<$-3.05 & 0.22 & $<$-9.76 \\ 
 Hn18$^\dag$ & M4 & 3270 & 0.8 & 0.11 & $<$-3.05 & 0.26 & $<$-9.85 \\ 
 Hn21W & M4.5 & 3200 & 2.2 & 0.12 & -2.40 & 0.22 & -9.08 \\ 
 \hline 
\end{tabular} 
\tablefoot{$^\dag$ Non-accreting objects according to the H$\alpha$ line width at 10\% of the peak and H$\alpha$ line equivalent width. All stellar parameters have been derived using the \citet{Baraffe98} evolutionary models apart from objects with a $^*$ symbol, for which the \citet{Siess00} models were used.} 
\end{center} 
\end{table*}  


\begin{table*}  
\begin{center} 
\footnotesize 
\caption{\label{tab::macc_others} Derived stellar and accretion parameters for the other targets in this study } 
\begin{tabular}{l| ccc| cc | cc } 
\hline \hline 
Object &  SpT & T$_{\rm eff}$ & A$_V$ & \lstar & log\lacc & \mstar & log\macc \\  
      &     \hbox{} & [K] & [mag] & [\lsun] & [\lsun] & [\msun] & [\msun/yr]   \\       
\hline 
\multicolumn{8}{c}{Taurus}\\
FN Tau & M3 & 3415 & 1.1 & 0.89 & -1.77 & 0.33 & -8.25 \\ 
 V409 tau & M0 & 3850 & 1.9 & 0.26 & -1.72 & 0.85 & -8.99 \\ 
 IQ Tau & M0 & 3850 & 1.7 & 0.75 & -1.09 & 0.99 & -8.19 \\ 
 DG Tau & K5 & 4350 & 2.2 & 1.55 & -0.44 & 1.40 & -7.65 \\ 
\hline 
\multicolumn{8}{c}{Chamaeleon~II}\\
 Hn24 & M0 & 3850 & 1.9 & 0.85 & -1.99 & 1.00 & -9.07 \\ 
 Sz50 & M4 & 3270 & 1.8 & 0.57 & -2.51 & 0.27 & -8.96 \\ 
 \hline 
\end{tabular} 
\end{center} 
\end{table*}

\subsection{Non-accreting targets in Chamaeleon~I}
All targets analyzed here have excess dust emission in Spitzer photometry and spectroscopy \citep[e.g.,][]{Luhman08,Manoj11}, indicating the presence of a disk in the system. Some of these, however, show no evidence of ongoing accretion from the lack of measurable excess in the Balmer continuum. In this section we discuss the properties of another proxy of accretion present in our spectra, the H$\alpha$ line profile.
The equivalent width (EW) and width at 10\% of the peak (W$_{10\%}$) of the H$\alpha$ line can be used to distinguish objects with ongoing accretion, although with discrepant results with respect to the Balmer continuum excess in some cases. 
According to \citet{WB03}, an object is defined as an accretor when W$_{10\%}>$270 km/s, and EW is higher than a threshold that depends on the SpT of the target. According to the \citet{WB03} criterion, seven targets in the sample are non-accreting: T4, ISO-ChaI 52, T33A, T45a, T54-A, Hn17, and Hn18. The fitting of the observed spectrum was also performed for these non-accreting YSOs. In most cases, the accretion excess in the Balmer continuum with respect to the photosphere is very small, within the noise of the spectra, or absolutely negligible, and this is compatible with the non-accreting status of these targets as suggested by the H$\alpha$ line profile. The value of \lacc \ derived in these cases is to be considered as an upper limit on the accretion rate. However, it seems that the excess in the Balmer continuum is non-negligible for ISO-ChaI 52, in particular, although we measure W$_{10\%}\sim$190 km/s and EW = -10.6 \AA. This object has been reported to accrete at a higher rate by \citet{Antoniucci11}, but their result is based on the flux of only few near-infrared emission lines, and they did not self-consistently derive the stellar parameters of the targets. It is also possible that the object is now accreting at a lower rate than in the past. For another non-accreting target, T4, \citet{Frasca15} reported a higher value of W$_{10\%}\sim$390 km/s than was measured from our spectra, W$_{10\%}\sim$230 km/s. This could be a variable object with a negligible accretion rate at the time of our observations.

\subsection{Targets in Chamaeleon~II and Taurus}
While the vast majority of the targets discussed in this work are located in the Chamaeleon~I region, four additional targets are located in Taurus and two are located in Chamaeleon~II. These were analyzed with the same method as described in Sect.~\ref{sect::method}, and their properties are reported in Table~\ref{tab::macc_others}. All these targets have a detectable accretion excess in the Balmer continuum and are also accretors according to the EW and W$_{10\%}$ of the H$\alpha$ line criteria. Two objects, namely FN-Tau and Sz50, have W$_{10\%}$ smaller than 270 km/s, but their EWs are larger than the threshold for their SpT. Only Hn24 presents a negligible excess in the Balmer continuum. This object has a very wide W$_{10\%}$ of more than 300 km/s. We report the accretion rate detected for this target, but the small excess and the small EW of the H$\alpha$ line indicate that it might be non-accreting.

\subsection{Comparison with other methods}

The stellar parameters (SpT, $A_V$, \lstar) for all the targets discussed here have previously been derived by several studies \citep[e.g.,][cf. Table~\ref{tab::lit}-\ref{tab::lit_others}]{Luhman07,Spezzi08,Daemgen13,Frasca15}. Accretion rate estimates are only available for some targets \citep[e.g.,][]{Antoniucci11,Antoniucci14,Frasca15}. Here we compare these results with our estimates.

Differences between our results and those in the literature are typically less than a spectral subclass in SpT and $\sim0.3$ mag in $A_V$, both within the uncertainties.
The largest differences in SpT occur for K-type stars, probably because our moderate-resolution blue spectra include better temperature diagnostics \citep[e.g.,][]{Covey07} than the low-resolution red spectra used in most previous studies of the region. The scarcity of templates in this work in the early-K spectral range may also contribute to this difference. Differences in $A_V$ estimates may be related to different methods, as \citet{Luhman07} used infra-red photometry and a different $R_V$ to derive extinction, while here the whole optical X-Shooter spectrum was used and $A_V$ is derived together with SpT and \lacc.   

The results derived with the method used here were also checked against other methods to derive SpT and \lstar \ from the observed spectra. In particular, the SpT determined here are also compared with those obtained using the spectral indices by \citetalias[][]{Herczeg14}, by \citet{Riddick07}, and the TiO index by \citet{Jeffries07}. In general, the agreement is very good with only a few exceptions in cases where different indices would lead to different results. Similarly, \lstar \ derived here agrees well, with differences in all the cases but one smaller than a factor $\sim$1.6, with those obtained using the bolometric correction by \citetalias{Herczeg14} after removing veiling.

\begin{figure}[!t]
\centering
\includegraphics[width=0.5\textwidth]{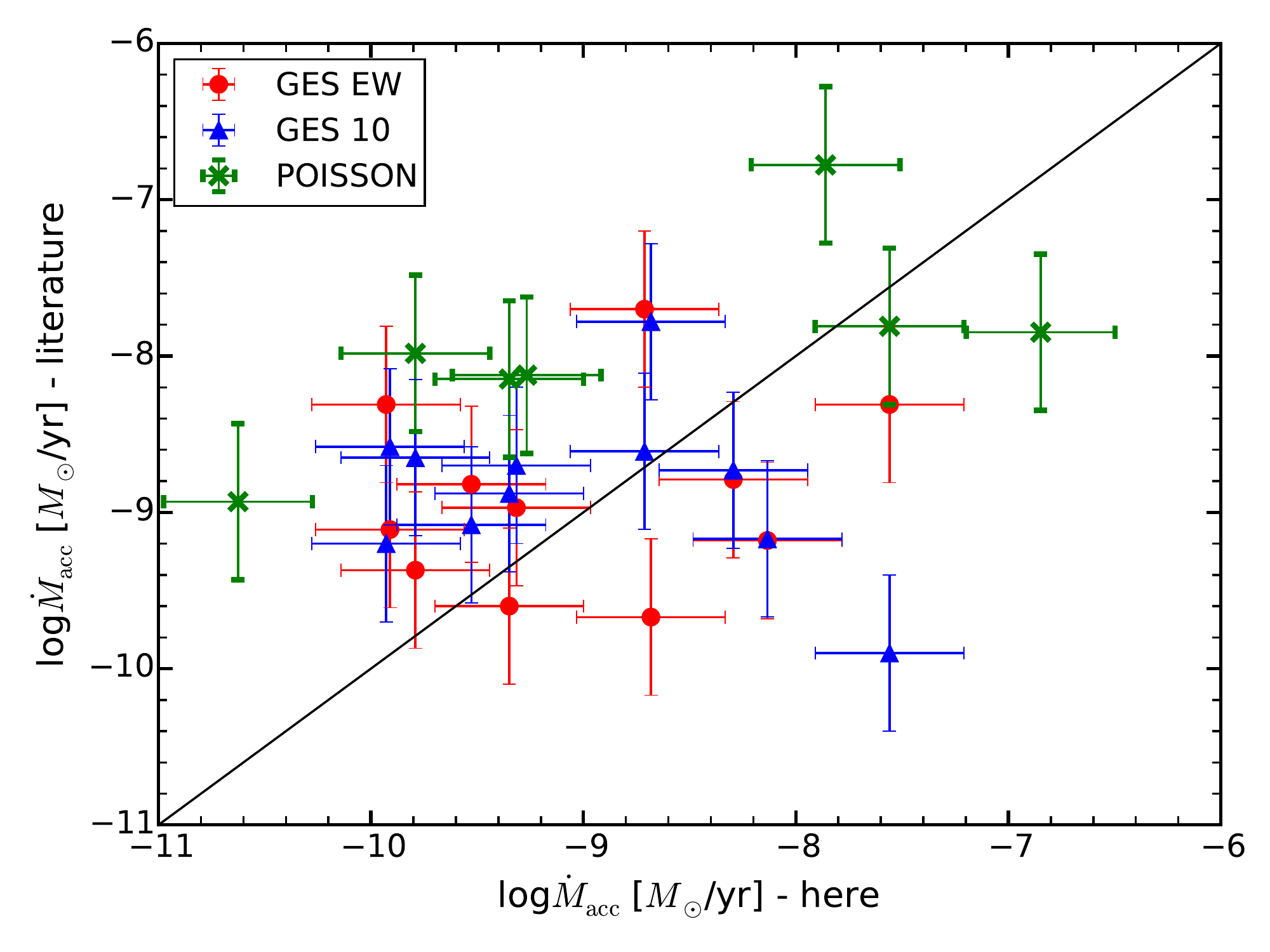}
\caption{Comparison of \macc \ derived here for targets in Chamaeleon~I with those from the literature, namely the POISSON survey \citep{Antoniucci11,Antoniucci14} and the Gaia-ESO survey \citep[GES][]{Frasca15}. The latter are derived either from the equivalent width of the H$\alpha$ line (EW) or from the 10\% width of the same line (10). 
     \label{fig::macc_compare}}
\end{figure}

Finally, Fig.~\ref{fig::macc_compare} shows the comparison between the values of \macc \ for Chamaeleon~I targets derived here and those from the POISSON survey \citep{Antoniucci11,Antoniucci14} and the Gaia-ESO survey \citep[GES,][]{Frasca15}. The POISSON survey used infrared spectra and the luminosity of the Br$\gamma$ and Pa$\beta$ emission lines to derive \lacc, and obtained \macc \ assuming the stellar parameters for the targets from \citet{Luhman07}. These values of \macc \ are different from those derived here in most cases. The reason for the difference is the use of a single indirect accretion indicator, and also the relation between the luminosity of the Br$\gamma$ line and \lacc \ they used, which differs from the more recent one by \citet{Alcala14}. The estimates of \macc \ from the GES were all derived using the EW and 10\% width of the H$\alpha$ emission line. Although our accretion rates generally agree with those from GES, some large discrepancies for individual objects are present.
The H$\alpha$ line, and especially its 10\% width, is known not to be the best tracer of an accretion rate \citep[e.g.,][]{Costigan12}, although it is widely used because it is the brightest emission line in YSOs. This comparison presented here confirms that single values of \macc \ from this indicator alone should be considered with caution.

\section{Results}\label{sect::results}
The dependence of the accretion rates on the stellar parameters is presented here. The relation between \lacc \ and \lstar \ is first discussed for the Chamaelon~I sample and in comparison with other large X-Shooter sample of YSOs, then the \macc-\mstar \ relation is presented. The results presented here are to be considered as only representative for the range of \mstar \ that
is\ covered with high completeness by this survey, namely at \mstar$\gtrsim$0.5\msun.

\subsection{Accretion luminosity and stellar luminosity dependence}\label{sect::lacc_lstar}

\begin{figure}[!t]
\centering
\includegraphics[width=0.5\textwidth]{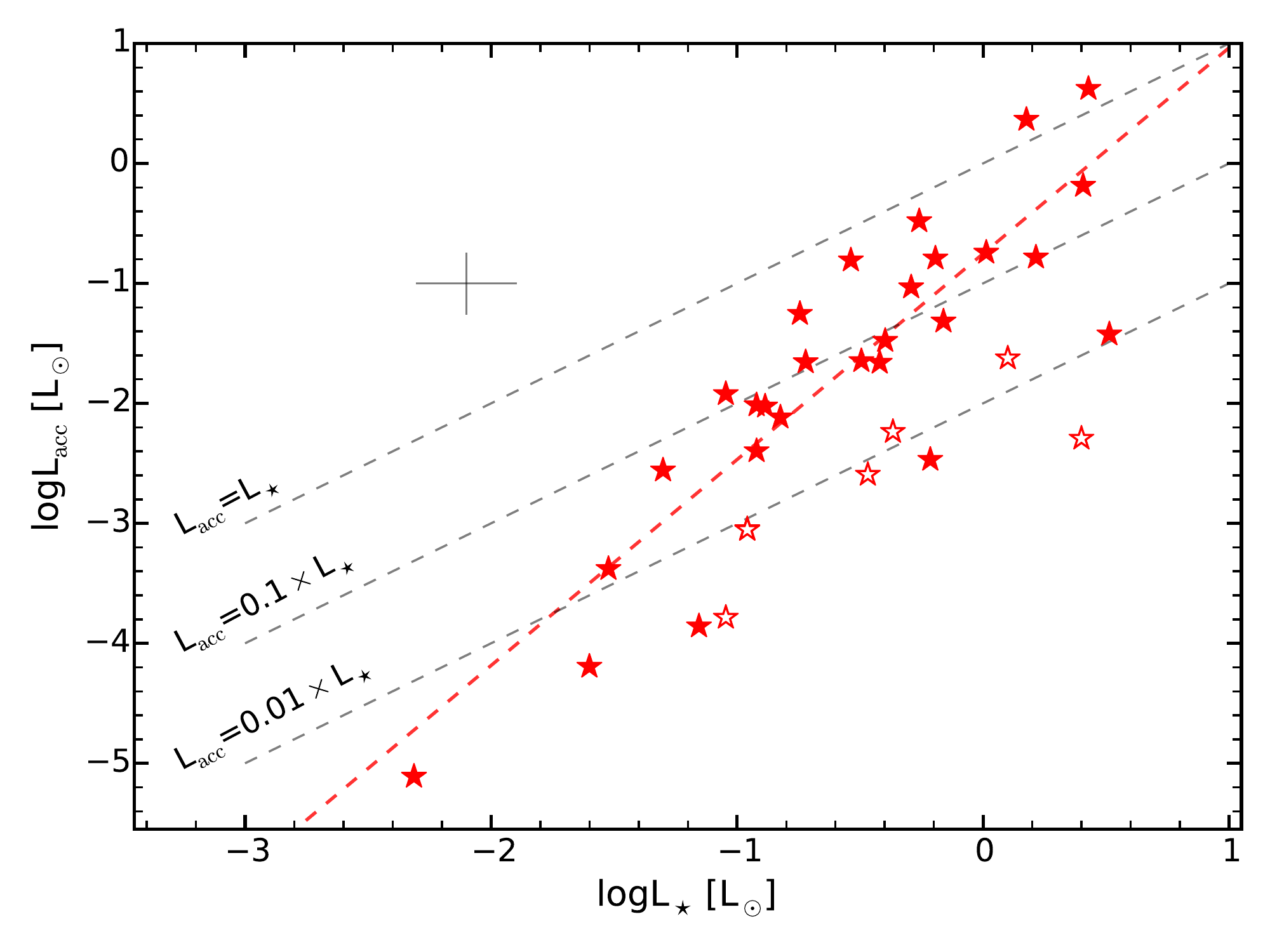}
\caption{Accretion luminosity as a function of stellar luminosity for the whole Chamaeleon~I sample discussed here (\textit{red stars}). Empty symbols denote non-accreting objects. Dashed lines are for different \lacc/\lstar \ ratios in decreasing steps from 1, to 0.1, to 0.01, as labeled. The best-fit relation from Eq.~(\ref{eq::lacc_lstar_chaI}) is overplotted with a dashed red line. Typical uncertainties are shown in the upper left corner of the plot.
     \label{fig::lacc_lstar_chaI}}
\end{figure}

The \lacc \ vs \lstar \ relation for the Chamaeleon~I sample is shown in Fig.~\ref{fig::lacc_lstar_chaI}. The targets mostly occupy the region between \lacc=\lstar \ and \lacc=0.01\lstar, although not uniformly at all stellar luminosity. Indeed, objects with \lstar$\gtrsim$0.1\lsun \ have \lacc/\lstar \ ratios ranging from $\sim$0.01 up to $\sim$1, or even more in two cases. These two strongly accreting targets both have \lstar$\sim$\lsun. In
contrast, objects with low stellar luminosity (\lstar$\lesssim$0.1\lsun) have 
 \lacc$\lesssim$0.1\lstar, and the objects with lowest luminosity have even \lacc$<$0.01\lstar. However, the sample discussed here comprises few objects in this range. Non-accreting objects, as expected, are located in the lower part of the plot, and they always have \lacc$\lesssim$0.01\lstar.

The positions of the targets on the \lacc-\lstar \ plane was fitted with the statistics package ASURV \citep{Feigelson} run in the IRAF\footnote{IRAF is distributed by National Optical Astronomy Observatories, which is operated by the Association of Universities for Research in Astronomy, Inc., under cooperative agreement with the National Science Foundation.} environment, considering the value of \lacc \ for the non-accreting objects as an upper limit. With the whole sample, we then find a best-fit relation of 
\begin{equation}
\log L_{\rm acc} = (1.58\pm 0.24)\cdot \log L_{\star} - (1.16\pm 0.21)
\end{equation}
with a standard deviation around the best fit of 0.89. If only accreting objects are analyzed, the best-fit line is
\begin{equation}\label{eq::lacc_lstar_chaI}
\log L_{\rm acc} = (1.72\pm 0.18)\cdot \log L_{\star} - (0.75\pm 0.16)
\end{equation}
with a standard deviation of 0.62. The latter relation is shown in Fig.~\ref{fig::lacc_lstar_chaI}.

\begin{figure}[!t]
\centering
\includegraphics[width=0.5\textwidth]{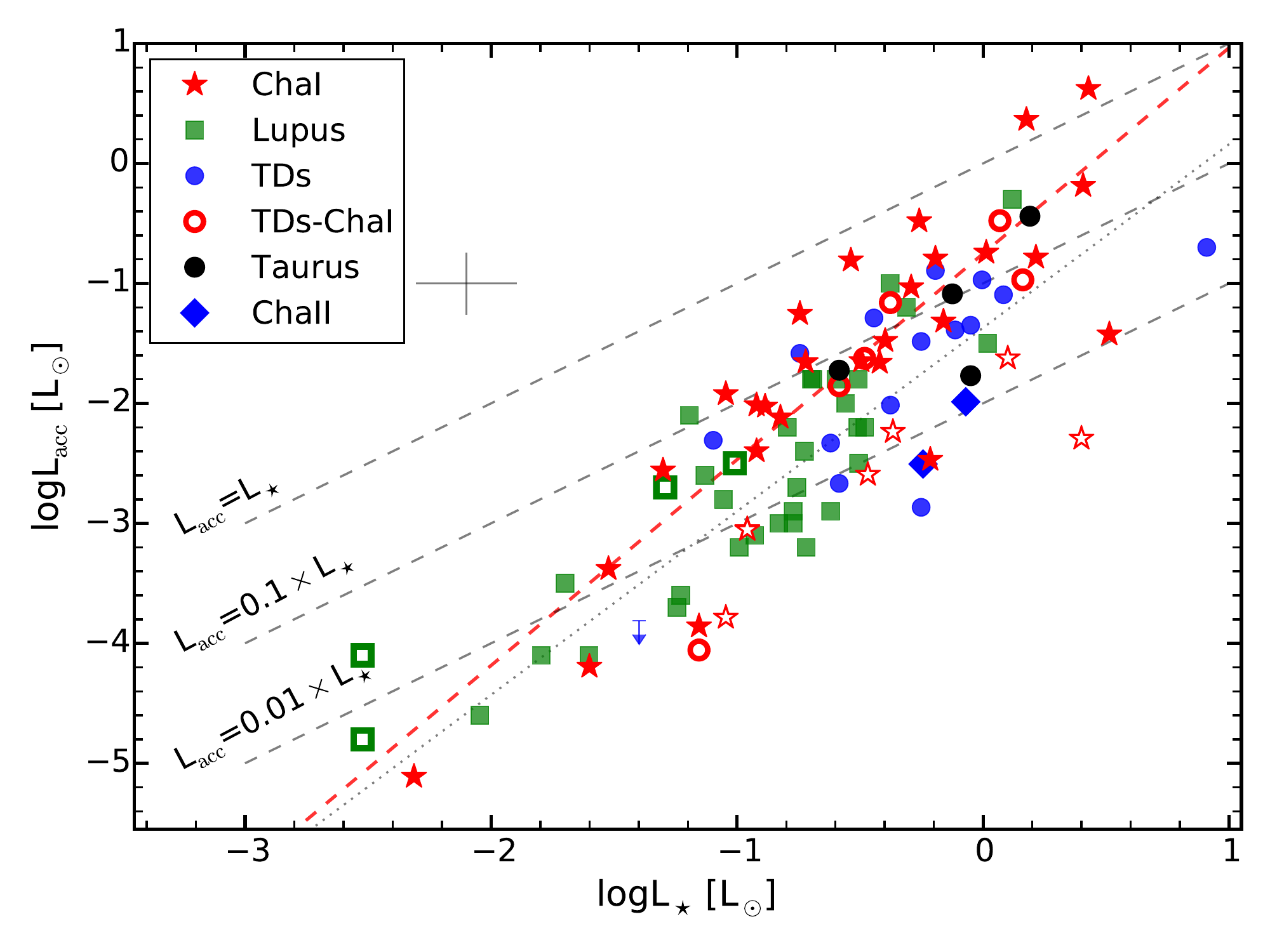}
\caption{Accretion luminosity as a function of stellar luminosity for the whole Chamaeleon~I sample of Class~II discussed here (red filled stars for accreting objects and empty stars for non-accreting objects) and for the Lupus sample of \citet{Alcala14}, reported with green squares, and with empty green squares for subluminous targets. Symbols for objects in other regions and at other evolutionary stages, such as transitional disks, are reported in the legend. In particular, transitional disks located in the Chamaeleon~I region are shown with red empty circles. Dashed lines are for different \lacc/\lstar \ ratios in decreasing steps from 1, to 0.1, to 0.01, as labeled.  The blue upper limit is a transitional disk.
     \label{fig::lacc_lstar}}
\end{figure}

In Fig.~\ref{fig::lacc_lstar} the values of \lacc \ vs \lstar \ for the Chamaeleon~I targets are compared with those in Taurus and Chamaeleon~II analyzed here, with the samples of Class~II YSOs in the Lupus clouds analyzed by \citet{Alcala14}, and with  transitional disks in various regions, including Chamaeleon~I, by \citet{Manara14}. These samples of X-Shooter spectra comprise the largest set of flux-calibrated broadband spectra of young stars used to simultaneously measure accretion, photospheric properties, and extinction. The distribution of stellar parameters for the objects in this sample is different, and comparisons can only be performed by extrapolating the results to other stellar parameters, for instance,  \lstar\  ranges.

The Taurus targets analyzed here follow the \lacc-\lstar \ distribution of the Lupus targets, while the Chamaeleon~II targets are located in the lower part of the distribution. With respect to the best fit of the sample of \citet{Alcala14} (shown in the figure as a black dotted line, from \citealt{Natta14}), the Chamaeleon~I targets follow a steeper slope of \lacc \ vs \lstar. However, the two values are compatible. Indeed, the slope of the best fit for this relation using the Chamaeleon~I data alone and excluding non-accreting objects is 1.7$\pm$0.2, while the slope derived from the Lupus sample is 1.5$\pm$0.2. 
In the Chamaeleon~I sample there are more strongly accreting YSOs with \lacc$\sim$\lstar than in the Lupus sample.
This difference in the fraction of strongly accreting YSOs in the Chamaeleon~I sample compared to that of \citet{Alcala14} may be the result of different selection criteria (see discussion in Sect.~\ref{sect::cha_vs_lup}). 
In any case, the sample analyzed here confirms that the relation between \lacc \ and \lstar \ has a slope significantly steeper than 1 and shows that there are several very strongly accreting YSOs in Chamaeleon~I. In contrast, no strongly accreting YSOs are present in the Lupus sample analyzed by \citet{Alcala14}. 

The distribution of the accreting transitional disks located in the Chamaeleon~I region shown in Fig.~\ref{fig::lacc_lstar} also seems to follow the best fit of the \lacc-\lstar \ relation of the Class~II YSOs in this region fairly well, with the only exception of CHXR22E, which is a non-accreting transitional disks that has been discussed by \citet{Manara14}. At the same time, the transitional disks from other regions are also mostly located in the same part of the \lacc-\lstar \ plane as the majority of Class~II YSOs, mostly following the locus of Lupus targets.

\subsection{Mass accretion rate and stellar mass dependence}\label{sect::macc_mstar}

\begin{figure}[!t]
\centering
\includegraphics[width=0.5\textwidth]{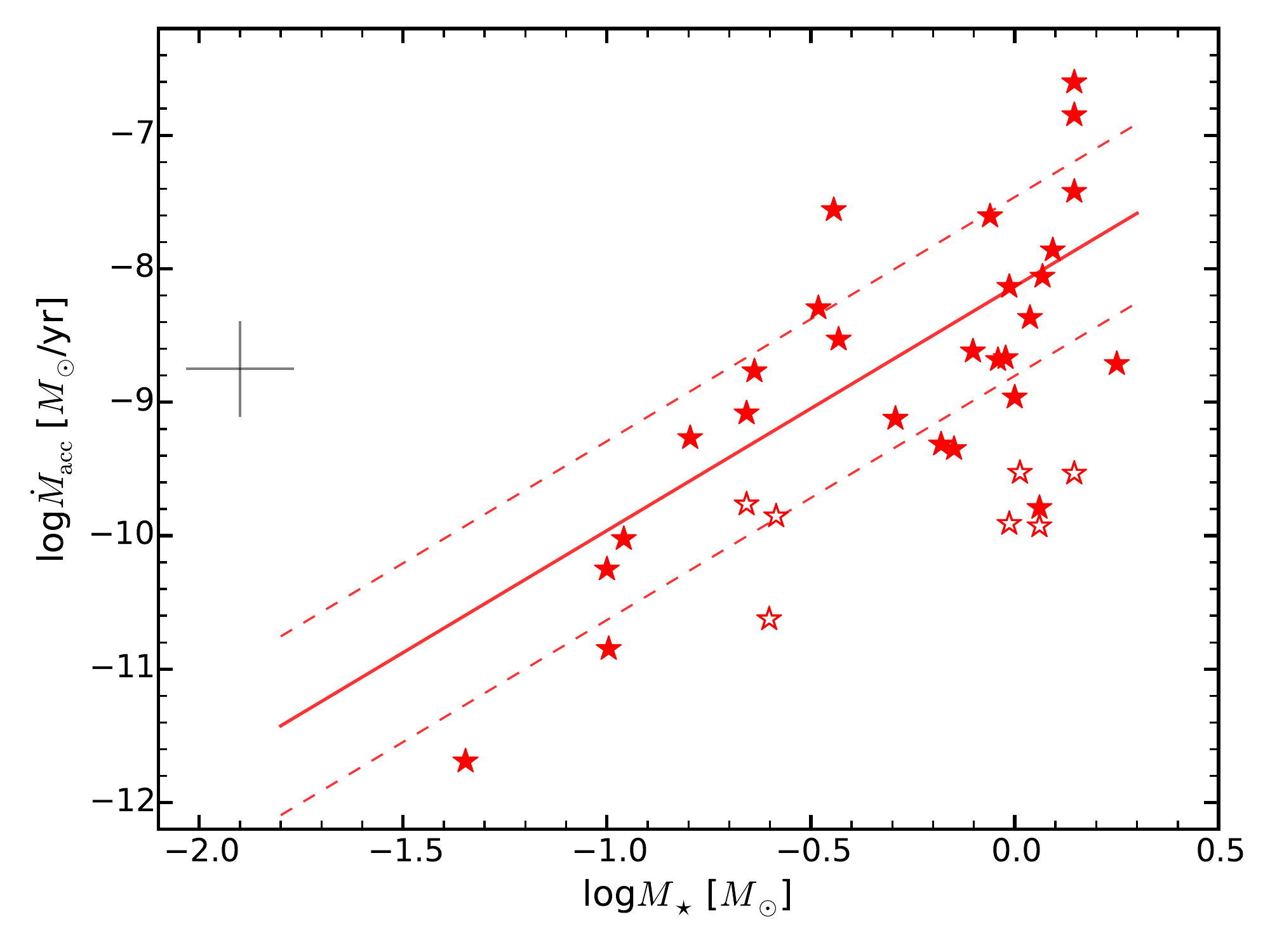}
\caption{Mass accretion rate as a function of the stellar mass for the whole Chamaeleon~I sample discussed here (\textit{red stars}). Empty symbols are used for non-accreting objects. The best-fit relation from Eq.~(\ref{eq::macc_mstar_chaI}) is overplotted with a red line, and the standard deviation from the best fit is shown with dashed lines. Typical uncertainties are shown in the upper left corner.
     \label{fig::macc_mstar_chaI}}
\end{figure}

The dependence of \macc \ on \mstar \ in the Chamaeleon~I sample is shown in Fig.~\ref{fig::macc_mstar_chaI}. All the non-accreting objects are located in the lower part of the distribution. Only ISO-ChaI 237 has a similar \macc \ as non-accreting objects, and it could also be borderline accretor. A tight correlation between \macc \ and \mstar \ is present, which is a well-known result in the literature. The best fit for this relation including non-accreting objects, again considered as upper limit on \macc, is
\begin{equation}
\log \dot{M}_{\rm acc} = (1.70\pm 0.44)\cdot \log M_{\star} - (8.65\pm 0.22)
,\end{equation}
with a standard deviation of 1.04. When excluding the non-accreting YSOs, the best fit is
\begin{equation}
\log \dot{M}_{\rm acc} = (1.94\pm 0.34)\cdot \log M_{\star} - (8.21\pm 0.17)
,\end{equation}
with a smaller standard deviation of 0.75.

However, several targets are outliers in this plot. Excluding Cha-H$\alpha$1, which is the only brown dwarf in the sample, and ISO-ChaI 237, whose position in the plot is in the same region as the non-accreting YSOs, the following best fit of the \macc-\mstar \ relation for the accretors in Chamaeleon~I is obtained:
\begin{equation}\label{eq::macc_mstar_chaI}
\log \dot{M}_{\rm acc} = (1.83\pm 0.35)\cdot \log M_{\star} - (8.13\pm 0.16)
\end{equation}
with a smaller standard deviation around the best fit of 0.67. Clearly, these two outliers influence the results substantially. This best fit and its standard deviation are shown in Fig.~\ref{fig::macc_mstar_chaI}. 

\begin{figure}[!t]
\centering
\includegraphics[width=0.5\textwidth]{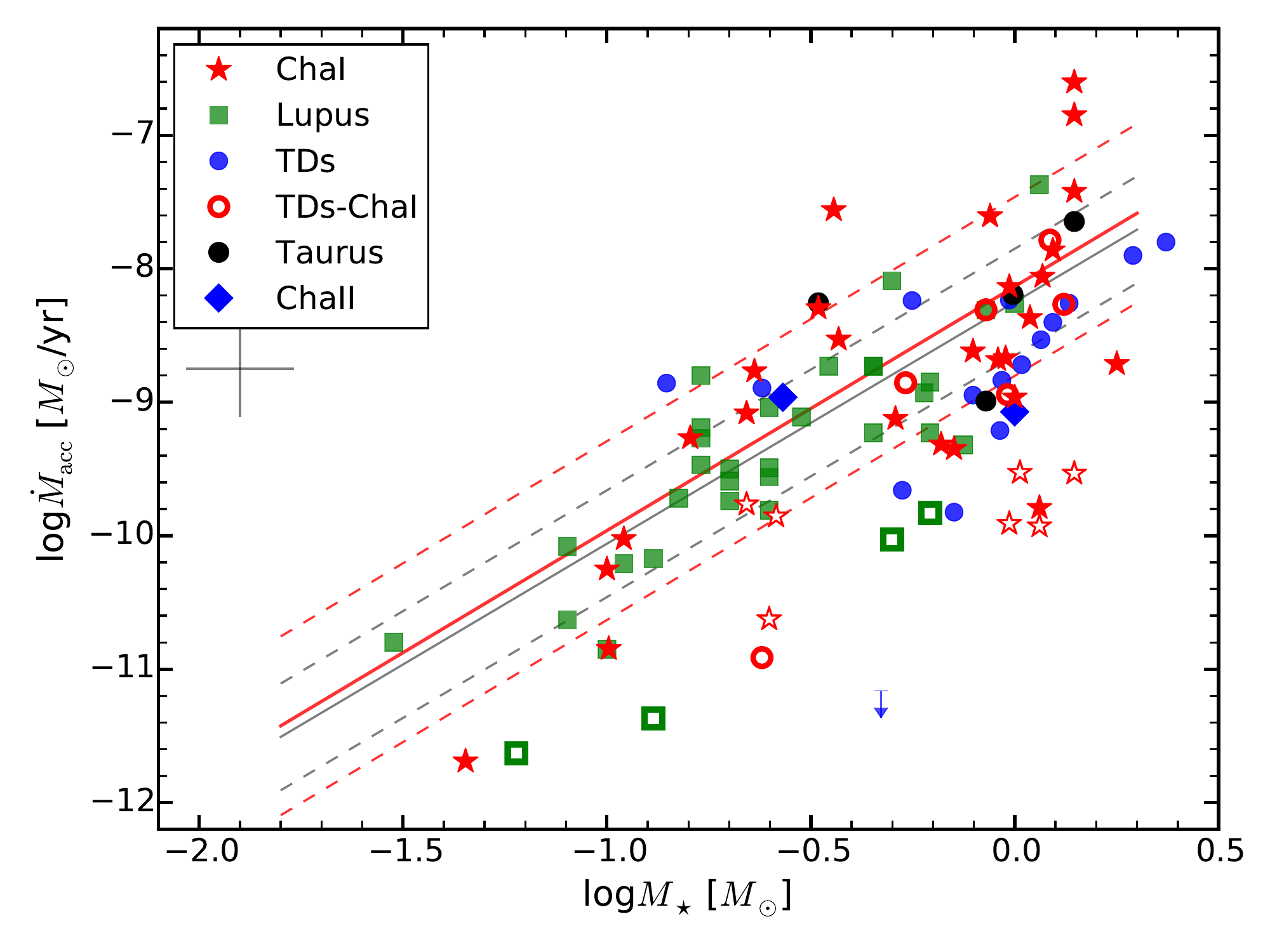}
\caption{Mass accretion rate as a function of mass for the whole Chamaeleon~I sample of Class~II discussed here (red filled stars for accreting objects and empty stars for non-accreting objects) and for the Lupus sample of \citet{Alcala14}, reported with green squares, and with empty green squares for subluminous targets. Symbols for objects in other regions and at other evolutionary stages, such as transitional disks, are reported in the legend. In particular, transitional disks located in the Chamaeleon~I region are shown with red empty circles. The blue upper limit is a transitional disk. The gray continuous line represents the linear fit of this relation for the Lupus sample \citep{Alcala14}, and the dashed lines represent the 1\,$\sigma$ deviation from the fit, while the red lines are the same for the Chamaeleon~I sample analyzed here.
     \label{fig::macc_mstar}}
\end{figure}


Figure~\ref{fig::macc_mstar} shows the \macc-\mstar \ relation for the same samples as in Fig.~\ref{fig::lacc_lstar}, that is, the Chamaeleon~I targets and those in Taurus and Chamaeleon~II analyzed here, the Lupus sample by \citet{Alcala14}, and the transitional disks by \citet{Manara14}. 
The few measurements in Taurus and Chamaeleon~II are consistent with the relations found in Chamaeleon~I and Lupus.
The slope of this relation is very similar between the Chamaeleon~I sample (1.83$\pm$0.35), when excluding the outliers and non-accreting YSOs, and the Lupus sample (1.81$\pm$0.20). 
The most relevant difference is then the larger spread around the best fit that is found in the Chamaeleon~I sample (0.67 dex) with respect to the spread of the Lupus targets (only 0.4 dex). In particular, there are some objects with \macc \ higher than that of objects with the same \mstar \ in Lupus at \mstar$\gtrsim$0.2\msun, but also several objects with lower \macc \ than the Lupus sample at \mstar$\sim$1 \msun.  It is also clear from Fig.~\ref{fig::macc_mstar} that there are several accreting Class~II YSOs in Chamaeleon~I with \macc \ similar to those of transitional disks, in particular similar to the transitional disks in Chamaeleon~I, but also to those in other regions.


\section{Discussion}\label{sect::discussion}

\subsection{Accretion rates in full and transitional disks}
The results presented in Sect.~\ref{sect::results} show that \lacc \ and \macc \ scale with the stellar properties, respectively \lstar \ and \mstar, with a high degree of correlation. In particular, the samples presented here and those from the literature include objects at different evolutionary stages of the disks, mainly Class~II YSOs, that is, objects surrounded by a full disk, and transitional disks. Although the comparison sample of transitional disks by \citet{Manara14} included here is small and biased toward known accretors, the comparison of \macc \ in transitional disks and full disks presented here shows no substantial differences. The values of \macc \ are similar for full and transitional disks even at \mstar$\sim$\msun. According to \citet{Najita15}, this is a result of the selected comparison sample. Indeed, they suggested that age differences between the targets might result in smaller differences in \macc, as \macc \ is expected to decrease with age because of the viscous evolution of the disks \citep[e.g.,][]{Hartmann98}. However, Fig.~\ref{fig::macc_mstar} shows that transitional disks located in Chamaeleon~I have \macc \  similar to full disks in the same star-forming region. 
A difference between the full- and transitional-disk Chamaeleon~I samples is that there are no transitional disks that are as strongly accreting as the strongest accretor from the sample of full disks. 
These results confirm that the sample of accreting transitional disks by \citet{Manara14} has similar accretion properties as full disks. However, no transitional disks in this sample have been identified as strong accretors. A possible difference between our analysis and the analysis by \citet{Najita15} may be the prevalence of strong accretors in their sample.
Again, this result is based on a small and biased sample of transitional disks and might not be valid for the whole class of objects.

\subsection{Empirical differences between Chamaeleon~I and Lupus}\label{sect::cha_vs_lup}

The results shown in Sects.~\ref{sect::lacc_lstar} and \ref{sect::macc_mstar} suggest that there are two main differences in the accretion rate properties of Chamaeleon~I and Lupus accreting YSOs: the Chamaeleon~I sample comprises more strongly accreting YSOs with \lacc$\sim$\lstar, and the spread of values of \macc \ at any given \mstar \ in the Chamaeleon~I sample is larger than the one found by \citet{Alcala14} for objects located in the Lupus clouds and is more compatible with the 0.8 dex found for Taurus members by \citet{Herczeg08}. In any case, this spread of \macc \ is still much smaller than was found in previous studies \citep[e.g.,][]{Muzerolle03,Mohanty05,Natta06,Manara12,Ercolano14}.

Before discussing the possible reasons for the difference in the accretion properties of objects in the two largest X-Shooter samples, it is important to remark that the Chamaeleon~I and the Lupus X-Shooter samples both include only $\sim$40-50\% of the targets surrounded by an optically thick disk (i.e., Class~II or transitional disks) in each of these regions. Therefore, selection effects are present in both samples, and some differences might be ascribed to this difference in the intrinsic stellar properties of the targets. In particular, the distribution of \mstar \ in the two samples is different: more than 70\% of the Lupus objects have \mstar$\lesssim$ 0.5~\msun, while 60\% of the Chamaeleon~I targets have \mstar$>$ 0.5~\msun. 
It is possible that the \macc-\mstar \ relation is tighter at lower \mstar, as recent results in low-mass stars and brown dwarfs (BDs) in $\rho$-Ophiuchus suggest \citep{Manara15}, and this would explain why there is a smaller spread of \macc \ in the Lupus sample. This means that it might be more plausible to have YSOs accreting with \lacc$\sim$\lstar \ at higher masses, which
would explain why more strongly accreting YSOs are present in the Chamaeleon~I sample. Both hypotheses can be tested only with more complete samples in various star-forming regions. 
The selection criteria of the two samples might likewise bias the results. The Lupus targets were selected mostly to be low-mass stars with low or negligible extinction, while the Chamaeleon~I targets sampled mostly objects with higher masses and bright, thus it might have been easier to include strongly accreting targets in the latter. 

In the following, we discuss some possible reasons for this difference in the context of accretion and disk evolution theories.

\subsection{Constraints on accretion and disk evolution theories}

The slope of the \macc-\mstar \ relation is a proxy for determining the main driver of the dispersal of protoplanetary disks \citep{Clarke06}. The facts that the slope is significantly higher than 1 and that there is no evidence of an upper locus of objects with slope unity in the \macc-\mstar \ plane disagrees with the expectations formulated by \citet{Hartmann06}. However, this locus of strongly accreting objects might not be observed in the samples analyzed here because the samples are incomplete. In the context of disk dispersal by photoevaporation, the slopes derived for both the Chamaeleon~I and Lupus samples are more compatible with X-ray photoevaporation models \citep{Ercolano14} than EUV photoevaporation models \citep{Clarke06}. If the larger spread of \macc \ values in Chamaeleon~I with respect to Lupus is driven by the different \mstar \ range of the two samples, this spread is indeed larger for higher mass stars than for BDs; this could be explained with the models by \citet{Alexander06}. They explain the \macc-\mstar \ relation as a consequence of initial conditions at formation of the disk. To reproduce the observed slope$\sim$2, they suggest that $t_\nu\propto M_\star^{-1}$, thus that higher mass stars evolve faster. However, observing this effect in a single star-forming region implies the presence of significant age spreads within that region and that the individual stellar ages may be reliably measured.  However, ages of individual young stars are unreliable at present \citep{SoderblomPPVI}.

In the context of viscous evolution theory \citep[e.g.,][]{Hartmann98}, a difference in age could explain the differences in the number of strong accretors and in the spread of \macc \ values between Chamaeleon~I and Lupus. The difference in age between the Lupus and Chamaeleon~I regions is probably smaller than 1-2 Myr, if there is any difference at all \citep[e.g.,][]{Luhman08,Alcala14}. Viscous evolution models \citep[e.g.,][]{Hartmann98} predict a typical decrease of \macc \ by $\sim$0.3 dex between 2 and 3 Myr for a typical low-mass YSO. This value is higher than the differences in the intercepts of the \macc-\mstar \ relations in the two samples and might explain some observed differences. However, the uncertainty on the age estimates and the possibility of non-coevality of the targets in each region lead us to consider this effect as a minor, if not negligible, explanation of the observed differences. 


As variability is often referred to as a possible explanation for the observed spread of \macc \ values, it is worth asking if it can play a role in explaining the differences between Lupus and Chamaeleon~I discussed here. \citet{Costigan12} have shown that the typical variability of \macc \ in Chamaeleon~I targets
is generally $\lesssim$0.4 dex. Independent studies in other star-forming regions also confirm this finding \citep[e.g.,][]{Venuti14}. Therefore, this effect can be relevant to explain some, even most, of the observed spread of values. If the main factor that explains the observed spread is variability, then this would suggest that the typical rate of accretion variability is lower for objects located in the Lupus cloud than for those in the Chamaeleon~I region.

We also consider whether the observed differences are due to the different environment in which the targets discussed here are located. \citet{Padoan05} suggested that the observed \macc \ are a consequence of Bondy-Hoyle accretion of material surrounding the YSO onto the central object. Such a description would imply a strong dependence of the values of \macc \ on the amount of pristine material in the region. However, the mean value of \macc \ is very similar in the Chamaeleon~I and Lupus regions, the largest difference is the scatter of the values. Therefore, this hypothesis could be tested by checking the amount of gas present in the immediate vicinity of individual YSOs and not at the level of the whole cloud. \citet{Dullemond06} also suggested that the spread of values in the \macc-\mstar \ relation is related to different initial conditions at the formation of disks, in particular different initial core rotation rates. If this is the case, then it might be possible that there has been a larger spread of core rotation rates in Chamaeleon I than in Lupus, or, on the other hand, that higher mass stars have a larger spread of this value at formation. 

Finally, it is also possible that the YSO population of Lupus is different from the populations in other regions. For example, \citet{Galli15} have found that the lifetimes of disks may be shorter in this region than the lifetimes for objects located in the Taurus star-forming region. Although this result relies on individual age estimates, it suggests that objects located in different environments might follow different evolutionary paths.

%

\section{Conclusions}\label{sect::conclusions}

We have presented the analysis of 34 X-Shooter spectra of YSOs in the Chamaeleon~I star-forming region, together with four targets in Taurus and two in Chamaeleon~II. We have derived their stellar and accretion parameters self-consistently from the same simultaneous spectra ranging from the UV excess, a direct accretion tracer, to the region at $\sim$700 nm, which is full of molecular features that are proxies of SpT of the targets. We distinguished between clearly accreting YSOs and those that do not fulfill the criteria to be considered accretors.

The dependence of accretion on the stellar parameters was tested with the logarithmic \lacc-\lstar \ and the \macc-\mstar \ relations, which both show a very good correlation when considering only the accreting targets. The slopes of these relations are 1.72$\pm$0.18 and 1.83$\pm$0.35, respectively. These values are compatible with those found in other star-forming regions, such as Lupus. More evolved objects, that is, transitional disks, also located in the Chamaeleon~I region follow the locus of full disks in these two relations very nicely as well. 

The largest discrepancy between our results and previous results with similar methods in Lupus is the larger spread of the \macc-\mstar \ relation around the best fit, and, in particular, the larger amount of strongly accreting YSOs we found. These differences might suggest that there are intrinsic differences in the evolution of protoplanetary disks in these two regions, but they might also be ascribed to the fact that these two samples cover only $\sim$50\% of the accreting YSOs in these regions and span a different range of stellar masses. The Chamaeleon~I sample analyzed here is mostly composed of YSO with masses closer to solar mass, while the Lupus sample comprises many lower mass stars. We might thus see a different behavior in a different stellar mass range. 

Our results suggest that the accretion properties of accreting YSOs located in the Chamaeleon~I  and in the Lupus regions might be different. This difference will be tested when complete X-Shooter samples in both star-forming regions, and possibly in more regions, will be available. At the same time, we await the results from current ALMA surveys of Chamaeleon~I and Lupus that will better determine the disk properties of these targets, as this could help us in understanding the differences between the measured accretion rates for objects with similar masses. Finally, with the data release of \textit{Gaia} we will have the possibility to refine the precision of our measurements with the newly determined distances to individual objects in these nearby star-forming regions.

\begin{acknowledgements}
      DF acknowledges support from the Italian Ministry of Science and Education (MIUR), project SIR (RBSI14ZRHR) and from the ESTEC Faculty Visiting Scientist Programme. We thank the referee for the valuable comments that helped to improve the quality of this paper. We acknowledge various discussions with Michael Rugel, Timo Prusti, Juan Alcal\'a. We thank Kevin Covey and Adam Kraus for help in preparing the proposal. This research made use of Astropy, a community-developed core Python package for Astronomy at http://www.astropy.org, of APLpy, an open-source plotting package for Python hosted at http://aplpy.github.com, and of the SIMBAD database, operated at CDS, Strasbourg, France 
\end{acknowledgements}


\appendix

\section{Additional data}

Additional information from the literature on the targets discussed here is reported in Tables~\ref{tab::lit}-\ref{tab::lit_others}.

\begin{table*}
\begin{center}
\footnotesize
\caption{\label{tab::lit} Data available in the literature for the targets in Chamaeleon~I included in this work }
\begin{tabular}{l|cc|ccc|c|c}  
\hline \hline
 Object/other name &  RA(2000)  & DEC(2000) &  SpT & A$_J$ &  Type & Notes & References \\ 
      &  h \, :m \, :s & $^\circ$ \, ' \, ''   &  \hbox{} & [mag] & \hbox{} & \hbox{} & \hbox{} \\         
\hline

T3 / SX Cha  & 10:55:59.7 & $-$77:24:39.91 & M0/M1.5 & 0.79 & II & Binary (2.21\arcsec) & 1,6,5 \\
T3 B / T3 E  & 10:56:00.0 & $-$77:24:40.6 & M3.5/M3 & 0.00 & ... & Binary (2.21\arcsec) & 1,5 \\
T4 / SY Cha  / Sz 3  & 10:56:30.4 & $-$77:11:39.31 & M0.5 & 0.23 & II & Accreting & 1,10,3 \\
TW Cha / T7 & 10:59:01.1 & $-$77:22:40.71 & K8 & 0.34 & II & Accreting, binary(spectroscopic?) & 1,6,3,9\\
CR Cha / T8 / Sz 6 & 10:59:06.9 &  $-$77:01:40.31 & K2 & 0.00 & II & ... & 1,6 \\
T12   / Sz10 / ISO$-$ChaI 10    &  11:02:55.0 & $-$77:21:50.81    & M4.5 & 0.32 & II   & Accreting & 1,2,3 \\
CT Cha A / T14 / Sz 11 & 11:04:09.1 & $-$76:27:19.38 & K5 & 0.45 & II & Accreting, binary (2.67\arcsec) & 1,2,3,5 \\
ISO$-$ChaI 52   &  11:04:42.6 & $-$77:41:57.13    & M4 & 0.36 & II  & Accreting & 1,2,3 \\ 
Hn 5  / GX Cha    &  11:06:41.8 & $-$76:35:48.98    & M4.5 & 0.32 & II & Accreting & 1,2,3 \\
T23     &  11:06:59.1 & $-$77:18:53.57    & M4.25 & 0.00 &  II & ... & 1,2 \\
T24 / UZ Cha / Sz 17  & 11:07:12.1 &  $-$76:32:23.23 & M0.5 & 0.63 & II & Accreting & 1,2,3 \\
Cha$-$H$\alpha-$1 / ISO$-$ChaI 95  &  11:07:16.7 & $-$77:35:53.29   & M7.75 & 0.00  & II & Accreting & 1,2,3   \\ 
Cha$-$H$\alpha-$9 / ISO$-$ChaI 98  &  11:07:18.6 & $-$77:32:51.66    & M5.5 & 0.45  & II & Accreting & 1,2,3 \\
Sz 22 / T29 / FK Cha / KG 28 & 11:07:57.9 & $-$77:38:44.93 & K6 & 1.24 & Flat & Accreting, triple (17.6\arcsec) & 1,2,8,11 \\
VW Cha / T31 / ISO$-$ChaI 123  & 11:08:01.5 & $-$77:42:28.85 & K8 & 0.72 & II & Accreting, binary(0.66\arcsec) & 1,2,3,4,5\\
ESO H$\alpha$ 562 / HO Cha / ISO$-$ChaI 126 & 11:08:02.9 & $-$77:38:42.59 & M1.25 & 1.35 & II & Accreting, binary (0.28\arcsec) & 1,2,3,5\\
T33 A / HP Cha / ISO$-$ChaI 135 & 11:08:15.1 & $-$77:33:53.16 & G7 & 0.85 & Flat & Accreting, triple (2.4\arcsec) & 1,2,3,4,5\\
T33 B  & ... & ... & K6/M0.5 & 0.52 & ... & Triple (2.4\arcsec), Binary (0.11\arcsec) & 1,5\\
ISO$-$ChaI 143 & 11:08:22.4 & $-$77:30:27.71 & M5 & 1.13 & II & Accreting, binary (18.16\arcsec) & 1,2,3,4,7 \\
Cha$-$H$\alpha-$6 / HQ Cha    &  11:08:39.5 & $-$77:34:16.67    & M5.75 & 0.38 & II & Accreting & 1,2,3,4\\
T38 / VY Cha / Sz 29 & 11:08:54.6 & $-$77:02:12.96 & M0.5 & 0.90 & II & Accreting & 1,2,3 \\
T40 / VZ Cha     &  11:09:23.8 & $-$76:23:20.76    &  K6 & 0.56 &  II & ... & 1,2 \\
T44 / WW Cha / ISO$-$ChaI 231 & 11:10:00.1 & $-$76:34:57.89 & K5 & 1.35 & II & Accreting & 1,2,3 \\
T45A / FN Cha  A & 11:10:04.7 & $-$76:35:45.22 & M1 & 0.54 & II & Binary (28.32\arcsec, ISO$-$ChaI 237) & 1,2,7 \\
ISO$-$ChaI 237 & 11:10:11.4 & $-$76:35:29.29 & K5.5 & 1.92 & II & Accreting, binary (28.32\arcsec, T45A) & 1,2,3,7 \\
T49 / XX Cha  & 11:11:39.6 & $-$76:20:15.25 & M2 & 0.34 & II & Accreting, binary (24.38\arcsec) & 1,2,3,7\\
CHX 18N & 11:11:46.3 & $-$76:20:09.21 & K6 & 0.20 & II & Accreting & 1,2,3 \\
T51 / CHX 20 / Sz 41  & 11:12:24.4 & $-$76:37:06.41 & K3.5/K7 & 0.00 & II & Binary (1.97\arcsec) & 1,2,5 \\
T51 B  & ...  & ... & M2.5 & 1.09 & ... & Binary (1.97\arcsec) & 5 \\
T52 / CV Cha   & 11:12:27.7 &  $-$76:44:22.30 & G9 & 0.42 &II & Binary (11.55\arcsec) & 1,2,8 \\
T54$-$A / CHX 22 & 11:12:42.7 & -77:22:23.05 & G8 & 0.60 & II & Binary (0.24\arcsec) & 1,2,5 \\
Hn17 & 11:12:48.6 & $-$76:47:06.67 & M4 & 0.09 & II & Accreting & 1,2,3 \\
Hn18 & 11:13:24.4 & $-$76:29:22.73 & M3.5 & 0.14 & II & Accreting & 1,2,3 \\
Hn21W  & 11:14:24.5 & $-$77:33:06.22 & M4 & 0.72 & II & Accreting, binary (5.48\arcsec) & 1,2,3,5 \\


\hline

\end{tabular}
\tablefoot{ Spectral types, extinction, disk classification, accretion indication, and binarity are adopted from the following studies: 1. \citet{Luhman07}; 2. \citet{Luhman08}; 3. \citet{Luhman04};  4. \citet{Costigan12}; 5. \citet{Daemgen13}; 6. \citet{Manoj11}; 7. \citet{Kraus07}; 8. \citet{Ghez97}; 9. \citet{Nguyen12};
10. \citet{Winston12}; 11. \citet{Schmidt13} . Complete name on SIMBAD for T\# is Ass Cha T 2$-$\# }

\end{center}
\end{table*}

\begin{table*}
\begin{center}
\footnotesize
\caption{\label{tab::lit_others} Data available in the literature for the targets in Taurus and Chamaeleon~II included in this work }
\begin{tabular}{l|cc|ccc|c|c}  
\hline \hline
 Object/other name &  RA(2000)  & DEC(2000) &  SpT & A$_V$ &  Type & Notes & References \\ 
      &  h \, :m \, :s & $^\circ$ \, ' \, ''   &  \hbox{} & [mag] & \hbox{} & \hbox{} & \hbox{} \\         
\hline

\multicolumn{8}{c}{Taurus (d=140 pc)}\\

FN Tau & 04:14:14.590 & +28:27:58.06 & M4.5 & 0.35 & II & d=131pc for HH14 & 1,2 \\
V409 tau & 04:18:10.785 & +25:19:57.39 & M0.6 & 1.00 & ... & d=131pc for HH14 & 1 \\
IQ Tau & 04:29:51.563 & +26:06:44.90 & M1.1 & 0.85 & II & ... & 1,2 \\
DG Tau & 04:27:04.698 & +26:06:16.31 & K7 & 1.60 & II & ... & 1,2 \\

\hline
\multicolumn{8}{c}{Chamaeleon~II (d=178 pc)}\\
Hn24 & 13:04:55.75 & $-$77:39:49.5 & M0 & 2.76 & II & TD? & 3,4 \\
Sz50 & 13:00:55.32 &  $-$77:10:22.2 & M3 & 3.78 & II & ... & 3,4 \\


\hline

\end{tabular}
\tablefoot{ Spectral types, extinction, disk classification, accretion indication, and binarity are from the following studies: 1. \citet{Herczeg14}; 2. \citet{Furlan11}; 3. \citet{Spezzi08}; 4. \citet{Alcala08}  }

\end{center}
\end{table*}

\section{Comments on individual targets}
\label{sect::ind_targ}

As discussed in Sect.~\ref{sect::method}, stellar and accretion properties of the targets discussed here were determined following the procedure described by \citet{Manara13b}. In a few cases, however, the quality of the spectra was not good enough for this procedure to work, or the properties of the target did not lead to satisfyingly good fits. In these cases slightly different procedures were adopted. Additional points were included in the fit of some targets, in particular at $\sim$400 nm and between $\sim$500 nm and 600 nm. In two cases, namely when fitting Cha H$\alpha$1 and Cha H$\alpha$9, the procedure was totally different. These cases are described in the following.

\subsection{Chamaeleon~I targets}

\textit{T3}: this target has a best fit with templates with SpT K7. However, the derived \lstar \ is low and results in a position on the HRD slightly below the 30 Myr isochrone according to the \citet{Baraffe98} models. Even including additional points in the fit did not change the results. However, the fit is good, the SpT is compatible with that from the literature (M0) and that from the TiO index by \citet{Jeffries07} (K6) and is slightly later than the result with the indices by \citetalias{Herczeg14} (K4.2). Similarly, the $A_V$ estimated here agrees well with previous values in the literature. When using the bolometric correction by \citetalias{Herczeg14} to determine \lstar, this is only $\sim$1.2 times higher than the one derived from the best-fit template. This would make the object only slightly younger, thus still older than 20 Myr.

\textit{CR Cha}: This object was best fit with a K0 template. Previous works reported it as an object with SpT K2, but the fit with a K2 template is not good. The value of SpT derived using the indices by \citetalias{Herczeg14} is K1, but no templates with this SpT are available. Given that the position of this target on the HRD is outside the range of the evolutionary models by \citet{Baraffe98}, the stellar parameters were derived using evolutionary models by \citet{Siess00}.

\textit{CT Cha A}: the best fit of this target was obtained by
including additional points at $\sim$400 nm and between $\sim$500 nm and 600 nm. If these additional points were not included, the best fit would have been with a K2 template, but it would have been worse and leading to a significantly older age. However, the other stellar and accretion parameters would be compatible within the uncertainties. When comparing \lstar \ determined from the best fit with that obtained using the bolometric correction by \citetalias{Herczeg14}, the latter would lead to a value higher by a factor $\sim$1.8. The reason for this discrepancy is possibly the high veiling in this object due to very intense accretion.

\textit{Cha H$\alpha$1}: None of the photospheric templates available in this study has the same SpT as this target. Indeed, literature estimates and spectral indices lead to an M7-M8 SpT. This object is also known to have negligible extinction \citep{Luhman07}, and it seems not to accrete at a high rate from the emission lines present in the spectrum. Therefore, it is possible to adopt the SpT from the spectral indices for this target, assume $A_V$=0 mag, determine \lstar \ using the bolometric correction by \citetalias{Herczeg14}, and \lacc \ from the luminosity of the emission lines present in the spectrum and the relations by \citet{Alcala14} to convert these luminosity in \lacc. This was the procedure adopted here to derive the stellar parameters reported in Table~\ref{tab::macc} and used in this work. 

\textit{Cha H$\alpha$9}: The spectrum of this object has almost no flux in the whole UVB arm, thus the procedure adopted in this work for the other targets was not applicable. This target is probably accreting at a very low rate according to its H$\alpha$ line. Indeed, the EW$_{\rm H\alpha}$ is 20 \AA, while the 10\% width of the line is $\sim$200 km/s. The stellar parameters were thus derived in the following way. First, the SpT was determined using several spectral indices, which led to an estimate of SpT M5.5. Then, the VIS part of the spectrum was compared to a photospheric template with SpT M5.5 that was reddened with increasing values of $A_V$ in steps of 0.1 mag until a best match was found. The best agreement between the two spectra is with $A_V$=4.8 mag, which is higher than that reported in the literature. Finally, \lstar \ was determined using the bolometric correction by \citetalias{Herczeg14} and \lacc \ from the only emission line detected with high S/N in the spectrum, which is the H$\alpha$ line. As the width of this line is quite small, it should be noted that the \lacc \ derived here might be an upper limit of the real \lacc, as this object might be non-accreting. 

\textit{Sz22}: This object is a very complex multiple system. Only the close binary at the center \citep{Schmidt13} is included in the slit. This object is a confirmed member of Chamaeleon~I by \citet{Lopez-Marti13} from proper motion studies. The best fit was obtained by including all the additional points listed above and with a value of $A_V$ much lower than the one reported in the literature \citep{Luhman07, Schmidt13}. No signs of the companion were found in the overall shape of the spectrum. The derived age is older than 10 Myr, but still compatible with younger ages according to other evolutionary models.

\textit{ESO H$\alpha$ 562}: The spectrum of this target has a low S/N ($\sim$1) in the UVB arm until $\sim$400 nm. It is reported to be a close binary (0.28\arcsec \ separation) of two objects with the same SpT (M0.5, \citealt{Daemgen13}). Therefore, both objects have probably been included in the slit and are not resolved. The best fit was obtained with SpT M1 and a \lstar = 0.11 \lsun. The same \lstar \ is obtained when using the bolometric correction by \citetalias{Herczeg14}. This places the object low on the HRD, at an age of $\sim$24 Myr. The $A_V$=3.4 mag derived here is smaller than the $A_V>$4 mag reported by \citet{Daemgen13}, while the \lstar \ determined here is the same as that reported by \citet{Daemgen13} for component B. It is thus possible that in the VIS and UVB arm the spectrum of component B dominates, as \citet{Daemgen13} reported that component A has $A_V$=10.5 mag and it is as bright as component B in the near-infrared. The latter has only $A_V$=4.3 mag, according to \citet{Daemgen13}, thus at optical wavelengths component B is significantly brighter than component A.

\textit{T33-B}: The best fit was made with a template with SpT K0, but this led to a low \lstar \ implying an age$\sim$40 Myr according to the evolutionary models by \citet{Baraffe98}. A similar \lstar \ was determined when using the bolometric correction by \citetalias{Herczeg14}. It should be noted that this component of the system T33 is composed of two objects separated by 0.11\arcsec \ \citep{Daemgen13}, thus not resolved by our observations. From near-infrared spectra, \citet{Daemgen13} reported a SpT M0.5 for these targets. However, both the fit and the spectral indices lead to a much earlier SpT and to an $A_V$ similar to that of component A.

\textit{Cha H$\alpha$6}: the best fit for this target was with the photospheric template with SpT M6.5, but this was obtained from a spectrum with a low S/N in the UVB. However, the spectral indices also confirm that the SpT of this target is around M6. The value of \lstar \ determined from the fit places the object immediately on the 1 Myr isochrone of the evolutionary models by \citet{Baraffe98}, while the \lstar \ from the bolometric correction by \citetalias{Herczeg14} leads to a smaller \lstar. As reported earlier on, this difference might be ascribed to the SpT-Teff relation used by \citetalias{Herczeg14}, which differs from that by \citet{Luhman03}, which was used here, at this late stellar type.

\textit{T44}: This object is a very strong accretor almost totally veiled, and it is known to launch a jet \citep{Robberto12}. The results of the fit are not perfect, for example, the slope of the Balmer continuum is not well reproduced.  However, even including additional points in the fit the accretion parameters change within the uncertainties, while the stellar parameters, in particular the SpT, might change. 

\textit{T45a}: This is a non-accreting target, but the profile of the H$\alpha$ line is unusual for a non-accreting object. It seems rather narrow ($\sim$150-170 km/s) and weak, but has a strong self-absorption at almost zero velocity. The excess in the UV is almost zero, and the \macc \ very low, consistent with being non-accreting. 

\textit{ISO-ChaI 237}: The spectrum of this target has a low S/N in the UVB arm, thus the best fit is slightly uncertain, but leads to a SpT compatible with the values from spectral indices and from the literature. This best fit leads to a low \macc \ in the \macc-\mstar \ plane, in the region where non-accreting targets are located. It is difficult to determine if this object is non-accreting because its spectrum presents a broad H$\alpha$ line but with strong self-absorption and, thus, a small EW. 

\textit{T49}: This object is one with stronger accretion than
most in the sample analyzed here. The spectrum is highly veiled, but a best fit was found and confirmed, also including additional points.

\textit{T54-A}: This is a non-accreting target, composed of two components separated by 0.24\arcsec, thus not resolved by our observations \citep{Daemgen13}. Component A should, however, dominate the emission in this wavelength range. The SpT derived here agrees with that from the literature and from spectral indices by \citetalias{Herczeg14}.

\Online

\section{Best fit}

Available on online version.

\onecolumn

\section{Observation log}

\landscape

\longtab{1}{
\begin{longtable}{lrrlllllll}
\caption{\label{tab::log} Night Log}\\
\hline\hline
Star & RA(J2000) & DEC(J2000) & Date [UT] & \multicolumn{3}{c}{Exposure
  Time [NDITxDIT(s)]} & \multicolumn{3}{c}{Slit Width [\arcsec]} \\
& & & & UVB & VIS & NIR & UVB & VIS & NIR \\
\hline
\endfirsthead
\caption{continued.}\\
\hline\hline
Star & RA(J2000) & DEC(J2000) & Date [UT] & \multicolumn{3}{c}{Exposure
  Time [NDITxDIT(s)]} & \multicolumn{3}{c}{Slit Width [\arcsec]} \\
& & & & UVB & VIS & NIR & UVB & VIS & NIR \\
\hline
\endhead
\hline
\endfoot
               FN Tau   &  04:14:14.70 & +28:27:55.1  &  2010-01-18;01:04:05.765  & 2x50  & 2x38  & 2x10  &  1.0 &  0.4 & 0.4  \\
             V409 tau   &  04:18:10.85 & +25:19:55.3  &  2010-01-18;01:13:22.953  & 2x60  & 2x40  & 2x20  &  1.0 &  0.4 & 0.4  \\
               IQ Tau   &  04:29:51.62 & +26:06:43.0  &  2010-01-18;01:22:07.078  & 2x90  & 2x80  & 2x10  &  1.0 &  0.4 & 0.4  \\
               ISO-ChaI 52   &  11:04:43.33 & -77:41:55.2  &  2010-01-18;04:10:03.932  & 2x360 & 2x350 & 2x360 &  1.0 &  0.4 & 0.4  \\
             Cha H$\alpha$9   &  11:07:17.92 & -77:32:52.3  &  2010-01-18;04:37:32.755  & 2x525 & 2x540 & 2x270 &  1.0 &  0.4 & 0.4  \\
       Ass Cha T 2 40   &  11:09:23.13 & -76:23:21.0  &  2010-01-18;05:02:09.994  & 2x45  & 2x40  & 2x20  &  1.0 &  0.4 & 0.4  \\
       Ass Cha T 2 23   &  11:06:58.50 & -77:18:53.2  &  2010-01-18;05:09:38.933  & 2x120 & 2x110 & 2x120 &  1.0 &  0.4 & 0.4  \\
             Cha H$\alpha$1   &  11:07:16.21 & -77:35:53.3  &  2010-01-18;05:29:37.917  & 4x420 & 4x410 & 4x210 &  1.0 &  0.4 & 0.4  \\
                 Hn 5   &  11:06:41.09 & -76:35:47.1  &  2010-01-18;06:27:00.167  & 2x120 & 2x110 & 2x60  &  1.0 &  0.4 & 0.4  \\
             Cha H$\alpha$6   &  11:08:39.05 & -77:34:15.3  &  2010-01-18;06:48:11.388  & 4x540 & 4x550 & 4x270 &  0.8 &  0.4 & 0.4  \\
       Ass Cha T 2 12   &  11:02:54.79 & -77:21:49.5  &  2010-01-18;07:34:15.838  & 2x150 & 2x140 & 2x75  &  1.0 &  0.4 & 0.4  \\
                Hn18   &  11:14:25.36 & -77:33:04.6  &  2010-01-18;08:09:13.040  & 2x530 & 2x540 & 2x270 &  1.0 &  0.4 & 0.4  \\
               DG Tau   &  04:27:04.29 & +26:06:08.7  &  2010-01-19;01:49:03.987  & 4x240 & 4x250 & 4x15  &  1.3 &  1.2 & 1.2  \\
        Ass Cha T 2 4   &  10:55:58.62 & -77:24:40.6  &  2010-01-19;04:43:50.328  & 2x110 & 2x120 & 2x120 &  1.0 &  0.4 & 0.4  \\
              ISO-ChaI 237   &  11:10:10.70 & -76:35:29.3  &  2010-01-19;04:55:06.057  & 2x170 & 2x180 & 2x180 &  1.0 &  0.4 & 0.4  \\
      Ass Cha T 2 45a   &  11:10:03.76 & -76:35:43.8  &  2010-01-19;05:09:28.923  & 2x90  & 2x100 & 2x100 &  1.0 &  0.4 & 0.4  \\
       Ass Cha T 2 51   &  11:12:25.16 & -76:37:09.3  &  2010-01-19;05:39:42.811  & 2x300 & 2x285 & 2x60  &  1.0 &  0.4 & 0.4  \\
                Hn17   &  11:12:52.43 & -76:47:06.3  &  2010-01-19;06:09:01.955  & 2x220 & 2x240 & 2x240 &  1.0 &  0.4 & 0.4  \\
                Sz22   &  11:07:57.19 & -77:38:43.5  &  2010-01-19;06:25:40.422  & 2x210 & 2x210 & 2x30  &  1.0 &  0.4 & 0.4  \\
           ESO H$\alpha$562   &  11:08:02.20 & -77:38:41.5  &  2010-01-19;06:37:21.103  & 2x360 & 2x345 & 2x180 &  1.0 &  0.4 & 0.4  \\
       Ass Cha T 2 54   &  11:12:42.35 & -77:22:21.2  &  2010-01-19;07:26:01.448  & 2x170 & 2x180 & 2x90  &  1.0 &  0.4 & 0.4  \\
       Ass Cha T 2 49   &  11:11:39.22 & -76:20:13.4  &  2010-01-19;07:38:58.735  & 2x200 & 2x210 & 2x210 &  1.0 &  0.4 & 0.4  \\
                Hn18   &  13:04:56.23 & -77:39:47.4  &  2010-01-19;07:52:42.296  & 2x240 & 2x250 & 2x250 &  1.0 &  0.4 & 0.4  \\
               VW Cha   &  11:08:00.96 & -77:42:27.4  &  2010-01-19;08:12:33.218  & 2x170 & 2x180 & 2x45  &  1.0 &  0.4 & 0.4  \\
               TW Cha   &  10:59:00.78 & -77:22:39.7  &  2010-01-19;08:46:24.914  & 2x150 & 2x160 & 2x50  &  1.0 &  0.4 & 0.4  \\
       Ass Cha T 2 52   &  11:12:27.67 & -76:44:21.0  &  2010-01-19;09:00:30.337  & 2x90  & 2x100 & 2x20  &  1.0 &  0.4 & 0.4  \\
                Hn18   &  11:13:23.75 & -76:29:22.1  &  2010-01-20;04:37:16.890  & 2x310 & 2x300 & 2x150 &  1.0 &  0.4 & 0.4  \\
        Ass Cha T 2 4   &  10:56:29.59 & -77:11:39.2  &  2010-01-20;04:54:15.929  & 2x120 & 2x110 & 2x60  &  1.0 &  0.4 & 0.4  \\
             CT Cha A   &  11:04:08.14 & -76:27:19.0  &  2010-01-20;05:26:22.036  & 2x120 & 2x110 & 2x60  &  1.0 &  0.4 & 0.4  \\
       Ass Cha T 2 24   &  11:07:11.65 & -76:32:22.2  &  2010-01-20;05:45:38.927  & 2x170 & 2x175 & 2x85  &  1.0 &  0.4 & 0.4  \\
              ISO-ChaI 143   &  11:08:21.60 & -77:30:26.9  &  2010-01-20;06:02:44.956  & 2x800 & 2x780 & 2x270 &  1.0 &  0.4 & 0.4  \\
       Ass Cha T 2 33   &  11:08:14.55 & -77:33:54.4  &  2010-01-20;06:56:21.776  & 2x90  & 2x80  & 2x10  &  1.0 &  0.4 & 0.4  \\
                SZ50   &  13:00:54.86 & -77:10:21.3  &  2010-01-20;08:02:17.861  & 2x170 & 2x160 & 2x20  &  1.0 &  0.4 & 0.4  \\
       Ass Cha T 2 38   &  11:08:54.50 & -77:02:10.1  &  2010-01-20;08:32:19.916  & 2x180 & 2x170 & 2x45  &  1.0 &  0.4 & 0.4  \\
               CR Cha   &  10:59:06.73 & -77:01:37.0  &  2010-01-20;09:02:48.053  & 2x65  & 2x60  & 2x20  &  1.0 &  0.4 & 0.4  \\
               CHX18N   &  11:11:46.50 & -76:20:06.4  &  2010-01-20;09:12:35.963  & 2x60  & 2x50  & 2x5   &  1.0 &  0.4 & 0.4  \\
               CHX18N   &  11:11:46.23 & -76:20:06.9  &  2010-01-20;09:20:42.895  & 4x35  & 4x30  & 4x15  &  1.0 &  0.4 & 0.4  \\
       Ass Cha T 2 51   &  11:10:00.41 & -76:34:55.3  &  2010-01-20;09:31:34.791  & 4x30  & 4x25  & 4x5   &  1.0 &  0.4 & 0.4  \\
\end{longtable}
}

\end{document}